\documentclass[article,amsmath,11pt,amssymb,floatfix,superscriptaddress,
showpagenumber,nofootinbib]{revtex4}
\usepackage{graphicx}
\usepackage[colorlinks=true, linkcolor=blue, urlcolor=blue, citecolor=blue]{hyperref}
\usepackage{epsfig}
\usepackage{bm}
\usepackage{amsfonts}
\usepackage{booktabs}
\usepackage{xcolor}
\usepackage{caption,subcaption}
\usepackage{graphicx}
\usepackage{graphicx}
\usepackage{float}
\usepackage{tabularx}
\usepackage{academicons}
\usepackage{xcolor}
\usepackage{orcidlink}

\input epsf

\begin{document}
\title{Cosmological Dynamics of Exponential Quintessence Constrained by BAO, Cosmic Chronometers, and DES-SN5YR/Pantheon$^+$ Data}

\author{Sanjeeda Sultana\,\orcidlink{0009-0003-7313-5926}\,\footnote{Email: sanjeeda.sultana0401@gmail.com; sanjeeda.sultana1@s.amity.edu}}
\author{Surajit Chattopadhyay\,\orcidlink{0000-0002-5175-2873}\,\footnote{Communicating author: Email: schattopadhyay1@kol.amity.edu; surajitchatto@outlook.com}}
\affiliation{ Department of Mathematics, Amity University Kolkata,\\
Major Arterial Road, Action Area II, Rajarhat, Newtown, Kolkata 700135,
India.}

\date{\today}

\newpage

\begin{abstract}
We perform a comprehensive observational test of a canonical quintessence model driven by an exponential potential, motivated by its emergence in higher-dimensional theories, string-inspired scenarios, and modified gravity. Using a Markov Chain Monte Carlo framework, we constrain the model with the latest high-precision observational datasets—Cosmic Chronometers, Baryon Acoustic Oscillation, Pantheon$^+$, and DES\text{-}SN5YR Type~Ia Supernovae. The combined data significantly tighten the parameter bounds on $(H_0, \Omega_{m_0}, \eta_0, \gamma)$ and yield predictions for the Hubble parameter $H(z)$, the distance modulus $\mu(z)$, and the scaled comoving angular diameter distance that remain in excellent agreement with observations and closely follow the $\Lambda$CDM baseline. In addition, an information-theoretic model comparison based on the Akaike Information Criterion shows that the exponential quintessence model remains statistically comparable with the $\Lambda$CDM scenario, despite the penalty associated with its additional parameters. The model naturally reproduces the transition from matter domination to late-time acceleration, maintains total
equation of state $\omega_{\mathrm{tot}} > -1$ as expected for canonical scalar fields, and provides an age of the universe consistent with \textit{Planck}~2018. The Statefinder diagnostics show trajectories approaching the $\Lambda$CDM fixed point, while allowing small observable departures. Energy condition analysis confirms physical viability, with only the Strong Energy Condition violated at late times, as required for acceleration. Overall, our results show that quintessence with an exponential potential provides a stable, observationally consistent alternative to $\Lambda$CDM. Using the latest high-precision datasets, we obtain some of the most stringent constraints on this exponential-potential quintessence model, reaffirming its viability as a compelling dynamical explanation for dark energy.\\

\end{abstract}

\maketitle

\section{Introduction}
Modern cosmology is deeply engaged in understanding the nature of dark energy, the component responsible for the observed late-time accelerated expansion of the universe. The first compelling evidence for this accelerated phase emerged from Type Ia Supernovae (SNe Ia) observations \cite{riess1998observational,perlmutter1999measurements}, which demonstrated that the cosmic expansion is speeding up rather than slowing down. These results imply that the present universe is dominated by an unknown form of energy—dark energy—which contributes roughly $70\%$ of the total energy budget. The remaining portion consists of matter, with only about $4\%$ in the form of ordinary baryonic matter, and the rest attributed to non-luminous dark matter. This picture has been independently supported by several other cosmological probes, including measurements of the Cosmic Microwave Background (CMB) anisotropies \cite{tegmark1999cosmological,Spergel_2003_148,ade2014planck,ade2016planck} and Baryon Acoustic Oscillation (BAO) observations \cite{Eisenstein_2005_633,sollerman2009first,Tegmark_2004_69}.

Dark energy \cite{nojiri2006dark,nojiri2006dark1} is commonly described as a perfect fluid with a negative equation of state (EoS) parameter $\omega = P/\rho$. The simplest and most widely studied candidate for dark energy is the cosmological constant $\Lambda$ \cite{Peebles_2003_75,peebles2020principles}, which corresponds to a fixed EoS $\omega = -1$. Within the standard $\Lambda$CDM framework, this constant term provides an excellent fit to a broad set of cosmological observations. A frequently cited physical interpretation associates $\Lambda$ with the vacuum energy arising from zero-point quantum fluctuations \cite{sahni2000case}, expressed as $\Lambda = 8\pi G \rho_{\mathrm{vac}}/c^{2}$. However, theoretical estimates of $\rho_{\mathrm{vac}}$ exceed observational bounds by many orders of magnitude \cite{Weinberg_1989_61,quartin2008dark}, leading to the well-known ``cosmological constant problem.'' 

Beyond this discrepancy, the $\Lambda$CDM model faces additional conceptual and observational challenges, such as the cosmological coincidence problem \cite{velten2014aspects} and significant tensions between cosmological parameters inferred from low-redshift probes \cite{riess2021cosmic,Riess_2019_876,joudaki2020kids+} and those derived from high-redshift measurements \cite{aghanim2020planck,odintsov2021analyzing}. These issues have motivated extensive efforts toward constructing alternative explanations for cosmic acceleration. The proposed solutions include modified gravity theories, which alter the geometric sector of Einstein’s field equations \cite{lusso2019tension,clifton2012modified,koyama2016cosmological,silvestri2013practical,bamba2015inflationary}, as well as dynamical or non-standard dark energy models that modify the energy-momentum sector \cite{copeland2006dynamics,pan2019interacting,karwal2022chameleon,murgia2021early,krishnan2021running,Bamba_2012_342,niedermann2020resolving,alam2004case,peracaula2021running,antoniadis2007cosmological}.

A broad class of dynamical dark energy models interprets dark energy as a particular form of matter, such as quintessence \cite{Caldwell1998}, k-essence \cite{ArmendarizPicon2000}, or the Chaplygin gas \cite{Kamenshchik2001}. In quintessence scenarios, a canonical scalar field evolves slowly along a self-interaction potential, analogous in spirit to the mechanisms employed in early universe inflationary models \cite{zlatev1999quintessence}. Comprehensive reviews of different quintessence realizations are available in Refs.~\cite{tsujikawa2013quintessence,caldwell2005limits,chiba2002quintessence,brax2000robustness,jesus2008new,perrotta1999extended,chimento2000enlarged,affleck1985dynamical,nomura2000quintessence,kim2003quintessential,panda2011axions,Carroll_1998_81}. 

Another prominent scalar-field framework is k-essence, which differs from quintessence by incorporating a non-canonical kinetic term; these models have also been extensively investigated in the literature \cite{armendariz2001essentials,malquarti2003new,rendall2006dynamics,armendariz2005haloes}. A key feature of such scalar-field approaches is their ability, in certain formulations, to alleviate the fine-tuning and coincidence problems while still producing the observed late-time cosmic acceleration. Unlike the $\Lambda$CDM model, in which the EoS parameter is fixed at $\omega=-1$, scalar-field dark energy models predict a dynamical EoS, providing richer phenomenology and potentially offering a deeper understanding of cosmic acceleration.

The behavior of the EoS parameter $\omega$ plays a central role in differentiating various quintessence scenarios. Broadly, quintessence models can be grouped into two categories: (i) thawing models and (ii) freezing models. In thawing scenarios, the scalar field remains effectively frozen at early times due to strong Hubble friction, held away from the minimum of its potential. Only at late times does the field begin to evolve, causing the EoS to deviate from $\omega_\phi \simeq -1$ and to increase gradually, such that $\omega'_\phi > 0$. 

In contrast, freezing models describe a scalar field that is already rolling down its potential in the early universe but gradually slows as its energy density becomes dominant. As a result, the EoS asymptotically approaches a value close to $-1$ in the late universe, corresponding to $\omega'_\phi < 0$. These distinct dynamical behaviors provide a useful framework for classifying and constraining quintessence models.

Quintessence models have been extensively investigated as potential alternatives to overcome the limitations of the standard $\Lambda$CDM framework. Numerous forms of the scalar-field potential have been proposed and analyzed in the literature \cite{copeland1998exponential,alho2015global,hill1988pseudo,amendola2000coupled,amendola2014multifield,jarv2004phase,matos2009dynamics,fang2009exact,odintsov2015inflation,leon2023scalar}. For such models to be viable, it is essential that the underlying dynamical system admits a stable attractor solution at late times, ensuring that a wide range of initial conditions converge to the same asymptotic behavior. Without this property, the model would suffer from fine-tuning issues similar to those encountered in the concordance cosmology. Furthermore, the chosen potential must generate a late-time accelerated expansion consistent with observations.

To assess these requirements, dynamical system methods offer a powerful framework for examining the qualitative behavior of scalar-field cosmologies. This approach is frequently employed to determine the stability properties of cosmological models \cite{coley2003dynamical,bhagat2025accelerating,fadragas2014dynamical}, particularly those involving scalar fields \cite{boehmer2012dynamics,pradhan2025dynamics}. Stability of the critical (or fixed) points is typically evaluated using the linear stability theorem \cite{raushan2021linear,bhagat2024observational}, while the centre manifold theorem is applied when non-hyperbolic critical points arise \cite{wiggins2003introduction,carr2012applications}.

The current study derives its primary motivation from the paper by Copeland et al. \cite{copeland1998exponential}. The exponential potential has attracted significant interest because of its appearances in various higher-dimensional theories, string-motivated models, and modified gravity reconstructions. Some significant works dealing with exponential potential \cite{Heard2002,Shahalam2017,Russo2004,Neupane2004,Kehagias2004}. The models based on this type of potential often exhibit scaling behavior and can reproduce viable models for late-time acceleration. In the current study, we aim to explore the parameter space of this form of potential in light of the most recent and high-precision cosmological datasets. This is motivated by the significant improvements in constraints on the universe's expansion history, as provided by the updated CC measurements, BAO datasets, Pantheon$^+$, and DES-SN5YR SNe Ia datasets. Motivated by these developments, the current study aims to reconstruct and constrain a canonical quintessence scalar field model with an exponential potential using the latest observational data. Combining multiple datasets and implementing MCMC, the study aims to determine whether this scalar field model can provide a consistent and viable alternative to $\Lambda$CDM, and to assess how different combinations of datasets influence parameter estimation and the resulting cosmological dynamics. The rest of the paper is organized as follows: Section~\ref{sec2} presents the foundational framework of the model, detailing the background dynamics of a canonical quintessence field governed by an exponential potential. In Section~\ref{sec3}, we employ a Bayesian inference framework implemented through the Markov Chain Monte Carlo (MCMC) sampling technique to constrain the parameters of the model considered and assess its viability by utilizing the latest cosmological observations, specifically Cosmic Chronometers (CC) measurements, BAO dataset and two distinct datasets for the SNe Ia data i.e. the Pantheon$^+$ compilation and the five-year Dark Energy Survey (DES-SN5YR) datasets. Section~\ref{sec4} presents a detailed analysis of the evolution of the key cosmological parameters within the framework of the proposed model. The principal findings of this work are summarized in Section~\ref{sec5}.

\section{Background and Formulation}\label{sec2}
We study a quintessence field \cite{radhakrishnan2025scalar} coexisting with non-relativistic matter, which is modeled as a barotropic perfect fluid. Since our interest lies in the late-time evolution of the universe, the influence of radiation is extremely small and can therefore be ignored. Consequently, only the contributions from matter and the scalar field are taken into account. The corresponding action is expressed as follows:
\begin{equation}
S = \int d^{4}x \sqrt{-g} \left[ \frac{1}{2\kappa^{2}}R - \frac{1}{2} g^{\mu\nu} \partial_{\mu}\phi \, \partial_{\nu}\phi - V(\phi) \right] + S_{m},
\label{10:1}
\end{equation}
where $\kappa^2=8\pi G$, $R$ is the Ricci scalar, $g^{\mu\nu}$ and $g$ are the metric
and its determinant respectively, and $S_m$ is the matter action. In this formulation, $V(\phi)$ denotes the potential that governs the dynamics of the scalar field $\phi$, which evolves by rolling down towards its minimum. To investigate the behavior of quintessence, we consider a spatially flat Friedmann-Lema\^{i}tre-Robertson-Walker (FLRW) cosmological background, specified by the curvature parameter $k=0$. The assumption of a flat geometry is well supported by current cosmological observations, such as those from the Cosmic Microwave Background (CMB), which strongly indicate that the universe is nearly spatially flat. For the quintessence scalar field $\phi$, the pressure and energy density are expressed as 
\begin{equation}
    P_{\phi} = \frac{1}{2}\dot{\phi}^{2} - V(\phi), \quad 
\rho_{\phi} = \frac{1}{2}\dot{\phi}^{2} + V(\phi),
\label{10:2}
\end{equation}
where the overdot denotes a derivative with respect to the cosmic time $t$. Consequently, the EoS parameter takes the form
\begin{equation}
\omega_{\phi} = \frac{P_{\phi}}{\rho_{\phi}} 
= \frac{\tfrac{1}{2}\dot{\phi}^{2} - V(\phi)}{\tfrac{1}{2}\dot{\phi}^{2} + V(\phi)} .
\label{10:3}
\end{equation}
In this framework, it is assumed that dark matter and the scalar field do not interact with each other, 
and therefore, each component obeys its own conservation law. 
For the scalar field, the conservation equation is given by
\begin{equation}
\dot{\rho}_{\phi} + 3H(\rho_{\phi} + P_{\phi}) = 0, \label{eq:scalar_conservation}
\end{equation}
while for dark matter it takes the form
\begin{equation}
\dot{\rho}_{m} + 3H\rho_{m} = 0. \label{eq:matter_conservation}
\end{equation}
Here, the dark matter is considered to be pressureless, which corresponds to the 
EoS parameter $\omega_{m} = 0$. By substituting the expressions for $\rho_{\phi}$ and $P_{\phi}$ into the conservation equation 
\eqref{eq:scalar_conservation}, one obtains the Klein--Gordon equation
\begin{equation}
\ddot{\phi} + 3H\dot{\phi} + V_{\phi} = 0, \label{eq:klein_gordon}
\end{equation}
where $V_{\phi} = \frac{dV}{d\phi}$ represents the derivative of the potential with respect to 
the scalar field, and $H = \frac{\dot{a}}{a}$ is the Hubble parameter. Within this framework, the Friedmann equations can be written as
\begin{equation}
3H^{2} = \kappa^{2}\left( \frac{1}{2}\dot{\phi}^{2} + V(\phi) + \rho_{m} \right), 
\label{eq:friedmann1}
\end{equation}
and
\begin{equation}
2\dot{H} = -\kappa^{2}\left(\dot{\phi}^{2} + \rho_{m}\right). 
\label{eq:friedmann2}
\end{equation}
Expressed in terms of the redshift $z$, the Klein-Gordon equation reads
\begin{equation}
(1+z)^{2} H^{2}(z)\,\phi''(z)
\;+\; (1+z)^{2} H(z)\,H'(z)\,\phi'(z)
\;-\; 2(1+z)\,H^{2}(z)\,\phi'(z)
\;+\; V_{\phi}(\phi) \;=\; 0,
\label{eq:KG_redshift}
\end{equation}
where $'$ and $''$ denote the first and second derivative, respectively, with respect to $z$ and $V_{\phi}(\phi)\equiv \frac{dV}{d\phi}$.

\subsection{Exponential Potential Function}
The exponential potential has been widely studied in cosmology due to its natural emergence in higher-dimensional theories, string-inspired frameworks, and modified gravity reconstructions \cite{copeland1998exponential,Heard2002,Shahalam2017,Russo2004,Neupane2004,Kehagias2004}. Models based on this potential often display scaling behavior and can yield viable late-time accelerated expansion. In this work, we investigate a canonical quintessence field governed by an exponential potential, motivated by the availability of high-precision cosmological datasets. The potential function is defined as
\begin{equation}
V(\phi)=V_{0}e^{-\kappa\gamma(\phi-\phi_{0})}.
    \label{10:4}
\end{equation}
Here, $V_{0}$ and $\phi_{0}$ denote the present-epoch values of the potential and the scalar field, respectively; i.e., their values at redshift $z=0$. From Eq.~(\ref{10:4}), we have
\begin{equation}
V_{\phi}(\phi)=-\kappa \gamma V(\phi).
    \label{10:5}
\end{equation}
and hence Eq.~(\ref{eq:klein_gordon}) can be expressed as
\begin{equation}
\ddot{\phi}= -3H\dot{\phi}+\kappa \gamma V(\phi).
    \label{10:6}
\end{equation}
Since obtaining a closed-form solution of Eq.~(\ref{eq:friedmann2}) for a generic potential $V(\phi)$ is
nontrivial, we adopt a numerical approach. For this purpose, it is convenient to define the
following dimensionless variables from Eq.~(\ref{eq:friedmann1}):
\begin{equation}
\eta \equiv \frac{\kappa\,\dot{\phi}}{\sqrt{6}\,H_0}, 
\qquad
\zeta \equiv \frac{\kappa\,\sqrt{V(\phi)}}{\sqrt{3}\,H_0},
\qquad
\xi \equiv \frac{\kappa^{2}\rho_{m}}{3H_0^{2}},
\qquad
h \equiv \frac{H}{H_0},
\label{eq:dimless_vars}
\end{equation}
where $H_{0}$ represents the present value of the Hubble parameter, i.e., its value at redshift $z=0$. The dimensionless variables introduced above satisfy the relation
\begin{equation}
h^{2} = \eta^{2} + \zeta^{2} + \xi,
\label{eq:h_relation}
\end{equation}
which connects the dimensionless variables with the normalized Hubble parameter $h$.

We can express Eq.~(\ref{eq:friedmann2}) in terms of the dimensionless variables as
\begin{equation}
-\frac{2\dot{H}}{3H^2}=\frac{2\eta^2+\xi}{h^2}.
    \label{10:7}
\end{equation}
By differentiating the variables $\eta$, $\xi$, and $h$ with respect to the redshift $z$, 
and making use of Eqs.~(\ref{eq:matter_conservation}), (\ref{10:6}), (\ref{eq:h_relation}) and (\ref{10:7}), the system can be expressed as a set of three coupled differential equations:
\begin{equation}
\frac{d\eta}{dz}=\frac{1}{1+z} \left( 3\eta - \gamma \sqrt{\frac{3}{2}} \, 
\frac{h^{2} - \eta^{2} - \xi}{h} \right),
\label{10:8}
\end{equation}

\begin{equation}
\frac{d\xi}{dz}=\frac{3\xi}{1+z},
\label{10:9}
\end{equation}

\begin{equation}
\frac{dh}{dz}=\frac{3}{2(1+z)h} \left(2\eta^{2}+\xi\right).
\label{10:10}
\end{equation}
By solving these equations numerically, we obtain the evolution of $\eta$, $\xi$, and $h$ 
as functions of the redshift $z$. We can define the matter density parameter $\Omega_{m}$  as well as the energy density parameter associated with the field, $\Omega_{\phi}$, through Eq.~(\ref{eq:dimless_vars}) in terms of variables $\xi$ and $h$ as
\begin{equation}
    \Omega_{m}=\frac{\xi}{h^2},
    \qquad
    \Omega_{\phi}=1-\frac{\xi}{h^2}.
    \label{10:11}
\end{equation}
The above formulation provides a self-consistent dynamical framework for studying the cosmological evolution of the scalar field with an exponential potential. By numerically integrating the coupled system of Eqs.~(\ref{10:8})--(\ref{10:10}), it is possible to investigate the redshift dependence of the field and matter components through the evolution of the dimensionless variables $\eta$, $\xi$, and $h$ using initial conditions $\eta_0$, $\xi_0=\Omega_{m_0}$ and $h_0=1$. In the context of the model considered, this framework introduces four independent free parameters: the present-day Hubble constant $H_0$, the current matter density parameter $\Omega_{m_0}$, $\eta_0$, and the model parameter $\gamma$. Consequently, the evolution of the fractional density parameters $\Omega_{m}$ and $\Omega_{\phi}$, along with the total EoS parameter $\omega_{tot}$, can be examined in detail to assess whether the model can explain the late-time acceleration of the universe. This approach, therefore, enables a detailed study of the dynamical behavior of the exponential potential and its consistency with observational cosmology.

\section{Observational Constraints and Model Parameter Estimation Methodology}\label{sec3}
In this study, we utilized the latest cosmological observations to constrain the parameters of the model considered and assess its viability. The analysis incorporates CC \cite{moresco2012improved,moresco2015raising} measurements, as well as the BAO \cite{Giostri_2012_2012_027,Raichoor_2020_500,Hou_2020_500} dataset.  We examined two distinct datasets for the SNe Ia data: the Pantheon$^+$ \cite{brout2022pantheonplus,scolnic2022pantheonplus} compilation and the DES-SN5YR \cite{abbott2024dark,vincenzi2024dark} dataset. These observational datasets provide precise information on the expansion history of the universe, allowing for a robust estimation of the model parameters. Further methodological details are presented in the following subsections.

\subsection{Bayesian Estimation of Cosmological Parameters via MCMC}

To estimate the cosmological parameter space $\boldsymbol{\Theta} = (H_0, \Omega_{m_0}, \eta_0, \gamma)$, we employ a Bayesian inference framework \cite{foreman2013emcee,karamanis2021zeus} implemented through MCMC sampling techniques \cite{Lewis2002, foreman2013emcee}. This approach enables an efficient scan of the multidimensional parameter space and yields statistically reliable posterior distributions for the model parameters.

Within the Bayesian formalism, the posterior probability is defined as
\begin{equation}\label{10:13}
P(\boldsymbol{\theta} \mid D, I) = \frac{P(\boldsymbol{\theta} \mid I) \cdot P(D \mid \boldsymbol{\theta}, I)}{P(D \mid I)},
\end{equation}
where $P(\boldsymbol{\theta} \mid I)$ denotes the chosen priors on the parameters, $P(D \mid \boldsymbol{\theta}, I)$ corresponds to the likelihood of the observed data $D$ conditioned on the model, $\boldsymbol{\theta}$ denotes the set of free parameters of the model under consideration, and $P(D \mid I)$ is the Bayesian evidence, acting as a normalization factor.

The likelihood is represented using the chi-squared statistic:
\begin{equation}\label{10:14}
\mathcal{L}(\boldsymbol{\theta}) = \exp\left(-\frac{\chi^2(\boldsymbol{\theta})}{2}\right).
\end{equation}
This analysis is applied to four complementary observational probes: CC, BAO, Pantheon$^+$, and DES-SN5YR.

\subsubsection*{\textbf{CC Dataset}}

The cosmic chronometer method \cite{Moresco_2022_25} offers a direct and model-independent way to determine the Hubble parameter $H(z)$ by utilizing the differential ages of massive, passively evolving galaxies. The technique is based on the fundamental relation $H(z) = -\frac{1}{1+z} \frac{dz}{dt}$,
where $dz/dt$ is inferred from the age difference of galaxies located at slightly different redshifts. Early-type galaxies that formed at high redshift and evolved passively serve as reliable tracers for this purpose. Since the determination of $H(z)$ through this approach does not depend on any predefined cosmological model, it provides an independent and stringent probe of the universe's expansion history.

For the present analysis, we employ a set of 32 CC measurements \cite{Borghi_2022_928} spanning the redshift interval $0.07 \leq z \leq 1.965$, collected from multiple spectroscopic surveys. Although the data points are relatively sparse, they furnish essential constraints on $H(z)$ at intermediate redshifts and are reported with associated observational uncertainties. The chi-squared function used to evaluate the model against the CC dataset is given by
\begin{equation}\label{10:15}
\chi^2_{\mathrm{CC}}(\boldsymbol{\theta}) = \sum_{i=1}^{32} \frac{\left[H_{\text{th}}(z_i; \boldsymbol{\theta}) - H_{\text{obs}}(z_i)\right]^2}{\sigma_H^2(z_i)},
\end{equation}
where $H_{\mathrm{th}}$ and $H_{\mathrm{obs}}$ correspond to the theoretical and observed values of the Hubble parameter at each redshift $z_i$, respectively, and $\sigma_H(z_i)$ denotes the measurement uncertainty.

\subsubsection*{\textbf{BAO Dataset}}

The BAO dataset probes the characteristic oscillatory features imprinted in the matter distribution of the early universe \cite{Giostri_2012_2012_027}. These baryon acoustic oscillations leave a measurable signature on the large-scale structure \cite{Raichoor_2020_500,Hou_2020_500}, making BAO observations a powerful tool for constraining cosmological models. By analyzing galaxy clustering both perpendicular and parallel to the line of sight, BAO measurements independently determine the angular diameter distance $d_{A}(z)$ and the Hubble expansion rate $H(z)$, respectively. The comoving sound horizon at the photon–decoupling epoch, $r_{s}(z_\ast)$, is given by
\begin{equation}
r_{s}(z_\ast) = \frac{1}{\sqrt{3}} \int_{0}^{1/(1+z_\ast)} 
\frac{d\tilde{z}}{\tilde{z}^{\,2} H(\tilde{z}) \sqrt{1 + \tilde{z} \left( \frac{3\Omega_{b0}}{4\Omega_{\gamma0}} \right)}},
\end{equation}
where $\Omega_{b0}$ and $\Omega_{\gamma0}$ represent the present-day baryon and photon density parameters, respectively. The quantity $z_\ast = 1091$ corresponds to the photon-decoupling redshift as inferred from the WMAP7 observations \cite{Larson_2011_192}. The comoving angular diameter distance and the BAO dilation scale are expressed as
\begin{equation}
d_{A}(z_\ast) = \int_{0}^{z_\ast} \frac{dz}{H(z)}, 
\qquad 
d_{V}(z_{BAO}) = \left[ d_{A}^{2}(z)\,\frac{z}{H(z)} \right]^{1/3}.
\end{equation}
These quantities encapsulate the transverse and radial clustering information extracted from BAO measurements. To evaluate the consistency between the theoretical model and the observed BAO data, we employ the chi-squared statistic
\begin{equation}
\chi^{2}_{\mathrm{BAO}} = X^{\mathrm{T}} C^{-1} X,
\end{equation}
where \(C\) denotes the covariance matrix associated with the BAO dataset, and \(X\) is the vector representing the differences between the theoretical predictions and the corresponding observational measurements.

In cosmological analyses, it is often convenient to rewrite the field equations in terms of the redshift $z$, since redshift is directly measured in astrophysical observations, unlike cosmic time $t$. The scale factor $a$ is related to redshift through 
\begin{equation}
z = a^{-1} - 1,
\end{equation}
with the present value normalized to $a(0)=1$. Using this relation, the time derivative can be transformed into a redshift derivative as
\begin{equation}
\frac{d}{dt} = \frac{dz}{dt}\,\frac{d}{dz} = -(1+z)H(z)\frac{d}{dz},
\end{equation}
which enables all cosmological quantities to be re-expressed as functions of $z$. This transformation provides a direct bridge between theoretical predictions and observable quantities.

\subsubsection*{\textbf{Pantheon$^+$ Dataset}}

The Pantheon$^+$ compilation \cite{brout2022pantheon,scolnic2022pantheonplus,brout2022pantheonplus} represents the most extensive and systematically improved collection of SNe Ia currently available. It comprises 1701 light curves corresponding to 1550 distinct SNe Ia events and covers the redshift range $0.00122 \leq z \leq 2.2613$. The dataset integrates observations from both ground-based surveys and space-based programs, including those conducted with the Hubble Space Telescope (HST).

Compared to the original Pantheon sample, Pantheon$^+$ features refined photometric calibration, updated zero-point corrections, enhanced treatment of systematic uncertainties, and revised host-galaxy bias corrections. These improvements lead to more accurate luminosity distances and tighter constraints on the late-time expansion history.

As in standard SNe Ia analyses, the theoretical distance modulus is given by
\begin{equation}\label{10:17}
\mu_{\text{th}}(z; \boldsymbol{\theta}) = 5 \log_{10} \left[d_L(z; \boldsymbol{\theta})\right] + \mu_0.
\end{equation}
Here, $\mu_0$ acts as a nuisance parameter that encapsulates both the absolute magnitude and the Hubble constant. The luminosity distance $d_L(z)$ is obtained from:
\begin{equation}\label{10:18}
d_L(z) = (1 + z) \int_0^z \frac{dz'}{E(z')},
\end{equation}
where $E(z) = \frac{H(z)}{H_0}$ denotes the dimensionless Hubble parameter, and $z'$ represents the integration variable running from $0$ to $z$. The associated chi-squared estimator used to constrain the model parameters is
\begin{equation}\label{10:19}
\chi^2_{\text{SN}}(\boldsymbol{\theta}) = \sum_{i=1}^{1701} \frac{\left[\mu_{\text{th}}(z_i; \boldsymbol{\theta}) - \mu_{\text{obs}}(z_i)\right]^2}{\sigma^2_\mu(z_i)}.
\end{equation}
where $\mu_{\text{th}}$ and $\mu_{\text{obs}}$ correspond to the theoretical and observed distance moduli, respectively, and $\sigma_\mu$ represents the associated uncertainties.

\subsubsection*{\textbf{DES-SN5YR Dataset}}

The five-year SNe Ia sample from the Dark Energy Survey (DES-SN5YR) \cite{abbott2024dark,vincenzi2024dark} is one of the most recent and statistically powerful compilations of SNe Ia aimed at probing late-time cosmic acceleration. The DES compilation contains 1829 SNe Ia in total, including 1635 events from the survey itself and 194 additional low-redshift SNe Ia, covering the redshift interval $0.025 \leq z \leq 1.13$. The dataset is based on observations conducted over five seasons by the Dark Energy Camera (DECam) on the Blanco 4-meter telescope at Cerro Tololo Inter-American Observatory. It combines both spectroscopically confirmed SNe Ia and high-quality photometrically classified events, offering an extensive and homogeneous sample across a wide redshift range.

The survey strategy was designed to optimize cadence, depth, and sky coverage, enabling precise light curve measurements and improved control of observational systematics. The DES-SN5YR dataset includes updated calibrations, refined photometric pipelines, and host-galaxy corrections, thereby enhancing the robustness of cosmological parameter estimation.

The chi-squared estimator adopted to confront the model with the DES-SN5YR data is
\begin{equation}\label{10:20}
\chi^2_{\text{DES}}(\boldsymbol{\theta}) = \sum_{i=1}^{1829} \frac{\left[\mu_{\text{th}}(z_i; \boldsymbol{\theta}) - \mu_{\text{obs}}(z_i)\right]^2}{\sigma^2_\mu(z_i)}.
\end{equation}

\subsubsection*{\textbf{Joint Likelihood and MCMC Inference}}

When multiple observational datasets\cite{bhagat2025logarithmic} are employed simultaneously, the total chi-squared function is constructed as the sum of the individual contributions. In the case of CC and BAO, the combined statistical measure takes the form
\begin{equation}\label{10:21}
\chi^2_{\text{CC+BAO}}(\boldsymbol{\theta}) = \chi^2_{\text{CC}}(\boldsymbol{\theta}) + \chi^2_{\text{BAO}}(\boldsymbol{\theta}).
\end{equation}
For extended combinations including SNe Ia samples, the total chi-squared becomes
\begin{equation}\label{10:021}
\chi^2_{\text{CC+BAO+SN}}(\boldsymbol{\theta}) = \chi^2_{\text{CC}}(\boldsymbol{\theta}) + \chi^2_{\text{BAO}}(\boldsymbol{\theta}) + \chi^2_{\text{SN}}(\boldsymbol{\theta}),
\end{equation}
where $\chi^2_{\text{SN}}$ corresponds to either the Pantheon$^+$ compilation or the DES-SN5YR dataset, depending on the specific analysis.

The posterior distribution is explored using MCMC techniques \cite{bhagat2025tracing}, ensuring adequate convergence and coverage of the parameter space. An initial burn-in phase is discarded to remove sampling transients, after which the remaining chains are used to compute marginalized posteriors and credible intervals for the cosmological parameters, including $H_0$, $\Omega_{m_0}$, $\eta_0$, and $\gamma$. The resulting probability distributions yield robust best-fit estimates and confidence regions, enabling a statistically consistent assessment of the viability of the model across different combinations of observational data.

\subsubsection*{\textbf{Interpretation of the MCMC corner plots}}

\begin{figure}[ht!]
    \centering
    \includegraphics[width=16cm]{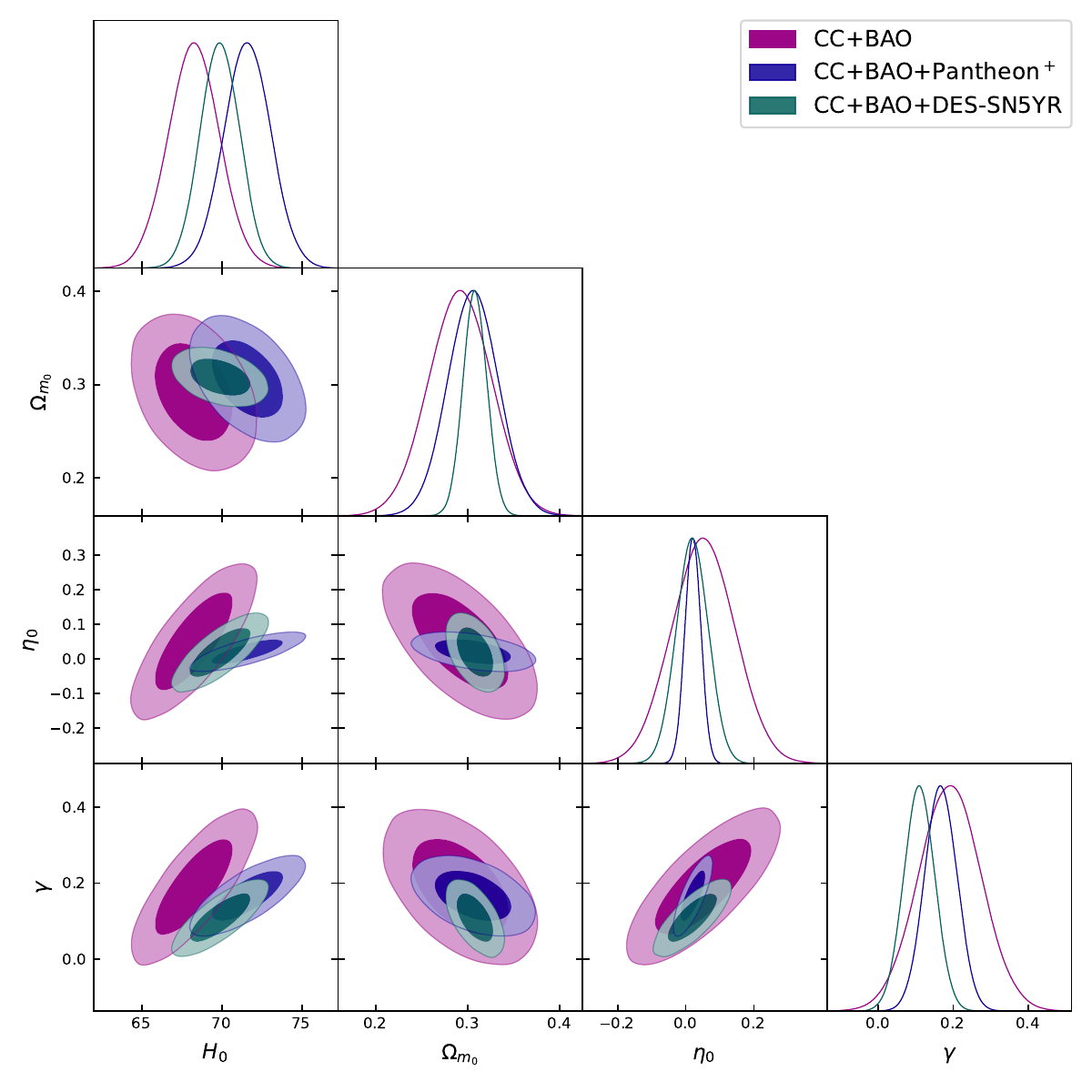}
  \caption{Corner plot showing the marginalized posterior distributions of confidence contours (68\% and 98\%) for the parameter space $\boldsymbol{\Theta} = (H_0, \Omega_{m_0}, \eta_0, \gamma)$ using CC+BAO, CC+BAO+Pantheon$^+$, and CC+BAO+DES-SN5YR dataset combinations.}
    \label{Fig1} 
\end{figure}

In Figure~\ref{Fig1}, we present the joint posterior distribution of the parameter space $\boldsymbol{\Theta} = (H_0, \Omega_{m_0}, \eta_0, \gamma)$ where $\eta_0$ is generated from the initial condition required to solve the differential equation Eq. (\ref{10:8}). This corner plot is based on three combinations of datasets: CC+BAO (magenta), CC+BAO+Pantheon$^+$ (blue), and CC+BAO+DES-SN5YR (green). The diagonal panels represent the marginalized posterior distributions. The panels above the diagonal panels exhibit 1$\sigma$ and 2$\sigma$ confidence contours for the pairs of parameters under consideration. It is evident from the contour plot that the inclusion of SNe Ia datasets, i.e., Pantheon$^+$ and DES-SN5YR, substantially tightens the confidence regions as compared to the CC+BAO alone. Hence, it can be inferred that the inclusion of SNe Ia data is enhancing the precision of the estimated cosmological parameters under study. The panel corresponding to the pair ($H_0$,$\Omega_0$) reflects the degeneracy between the expansion rate and the matter content of the universe. On the other hand, there exist moderate correlations for the pairs ($H_0$,$\eta_0$), ($H_0$,$\gamma$), and ($\eta_0$,$\gamma$). This implies that the combination of model parameters $H_0$ and $\gamma$ jointly influences the cosmological dynamics of the model under study. Furthermore, we observe a noticeable narrowing of the distributions of $\Omega_{m_0}$ and $h_0$ for the combined datasets. This suggests stronger convergence towards the best-fit values of the cosmological parameters. Another noteworthy observation is that although $\eta_0$ is not a model parameter, the mutual influence of $\eta_0$ and the model parameter $\gamma$ indicates suitable initial value selection while solving the differential equation Eq. (\ref{10:8}). Overall, the corner plot presented in Figure \ref{Fig1} suggests that the model under consideration yields well-constrained and consistent parameter values when tested against the latest observational data. Table~\ref{table:1} provides a comprehensive summary of the parameter constraints obtained for the reconstructed scalar field model based on three different combinations of observational datasets. To be more specific, we report in the table the best-fit values and corresponding confidence intervals for the four model parameters derived from CC+BAO, CC+BAO+Pantheon$^+$, and CC+BAO+DES-SN5YR datasets.

\begin{table}[hbt]
\renewcommand\arraystretch{1.5}
\centering 
\begin{tabular}{||c|c|c|c|c||} 
\hline\hline 
~~~Dataset~~~&~~~~~~~ $H_{0}$ ~~~~~~~& ~~~~~~~$\Omega_{m_0}$~~~~~~~ & ~~~~~~~~~~$\eta_0$~~~~~~~&~~~~~~~$\gamma$~~~~~~\\ [0.5ex]  
\hline\hline
CC+BAO & $68.24\pm 1.67$ &   $0.292^{+0.036}_{-0.032}$ & $0.052\pm0.093$ & $0.191 \pm 0.084$ \\[0.5ex]
\hline
CC+BAO+Pantheon$^+$ & $71.60^{+1.53}_{-1.49}$ &  $0.306\pm0.028$ & $0.021^{+0.023}_{-0.021}$ & $0.167\pm0.043$\\[0.5ex]
\hline
CC+BAO+DES-SN5YR & $69.870\pm 0.203$ &  $0.308 \pm 0.013$ & $0.019^{+0.045}_{-0.048}$ & $0.109^{+0.041}_{-0.043}$\\[0.5ex]
\hline \hline 
\end{tabular}
\caption{Constrained parameter values obtained for different datasets for the reconstructed Quintessence gravity model, obtained using three observational datasets: CC+BAO, CC + BAO+Pantheon$^+$, and CC +BAO+ DES-SN5YR.}
\label{table:1} 
\end{table}

\begin{figure}[htbp]
\centering
\includegraphics[width=1.1\textwidth]{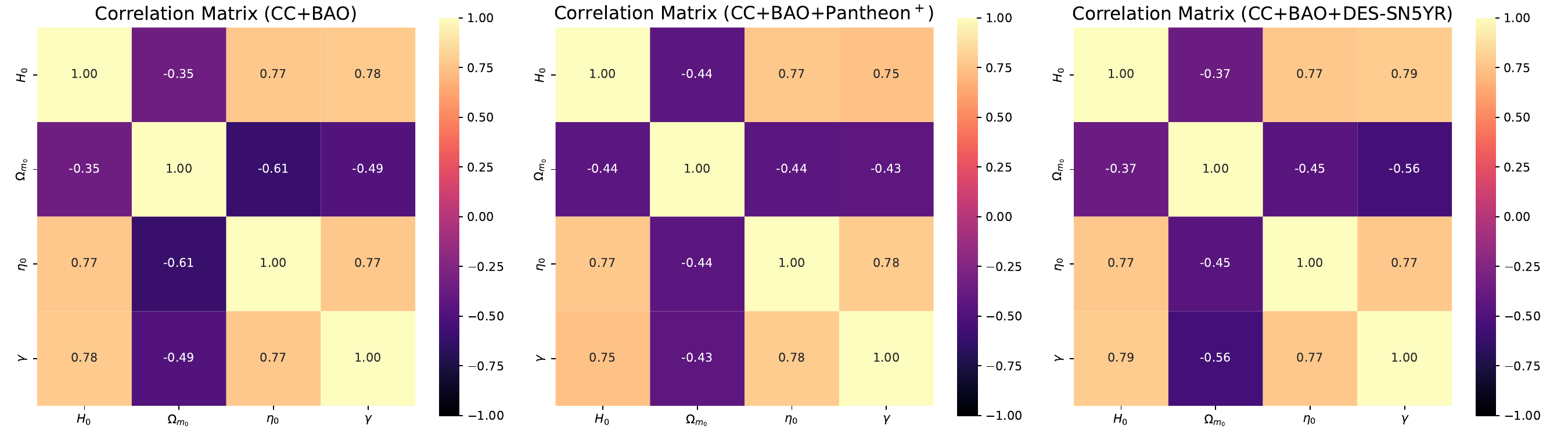}
\caption{Correlation matrices for the model parameters \( H_0 \), \( \Omega_{m_0} \), \( \eta_0 \), and \( \gamma \), obtained from the CC+BAO, CC+BAO+Pantheon$^+$, and CC+BAO+DES-SN5YR dataset combinations. \label{Fig6}}
\end{figure}

Figure~\ref{Fig6} illustrates the heat map representing correlation matrices for the parameters \( H_0 \), \( \Omega_{m_0} \), \( \eta_0 \), and \( \gamma \), obtained from the CC+BAO, CC+BAO+Pantheon$^+$, and CC+BAO+DES-SN5YR dataset combinations. This figure consists of three panels corresponding to three combinations of observational data. The intensity of color and shade represents the degree of association in both positive and negative aspects. It is known that the correlation value close to $+1$ signifies that if the true value of a parameter A is slightly higher, then the true value of the parameter B is also slightly higher. In case of a value close to $-1$, a higher A implies a lower B. In Figure~\ref{Fig6}, a negative correlation between $H_0$ and $\Omega_{m_0}$ is observed in all three panels. Both the current expansion rate and the amount of matter today influence the overall expansion rate $H(z)$. It can be interpreted from the figure that a slightly lower $H_0$ is compensated by the higher $\Omega_{m_0}$. This is the reason behind their significant negative correlation. A similar kind of association is observable for the pairs ($\Omega_{m_0}$,$\eta_0$) and ($\Omega_{m_0}$,$\gamma$). As a natural consequence, a strong positive correlation happens to occur for the pairs ($H_{0}$,$\eta_0$) and ($H_{0}$,$\gamma$). It is worth noting that, regardless of the parameter combinations, the correlational pattern remains remarkably similar across all observational datasets.

\subsubsection*{\textbf{Observational Constraints on $H(z)$ from Combined Datasets and Comparison with the $\Lambda$CDM Model}}

\begin{figure}[htbp]
\centering
\includegraphics[width=1\textwidth]{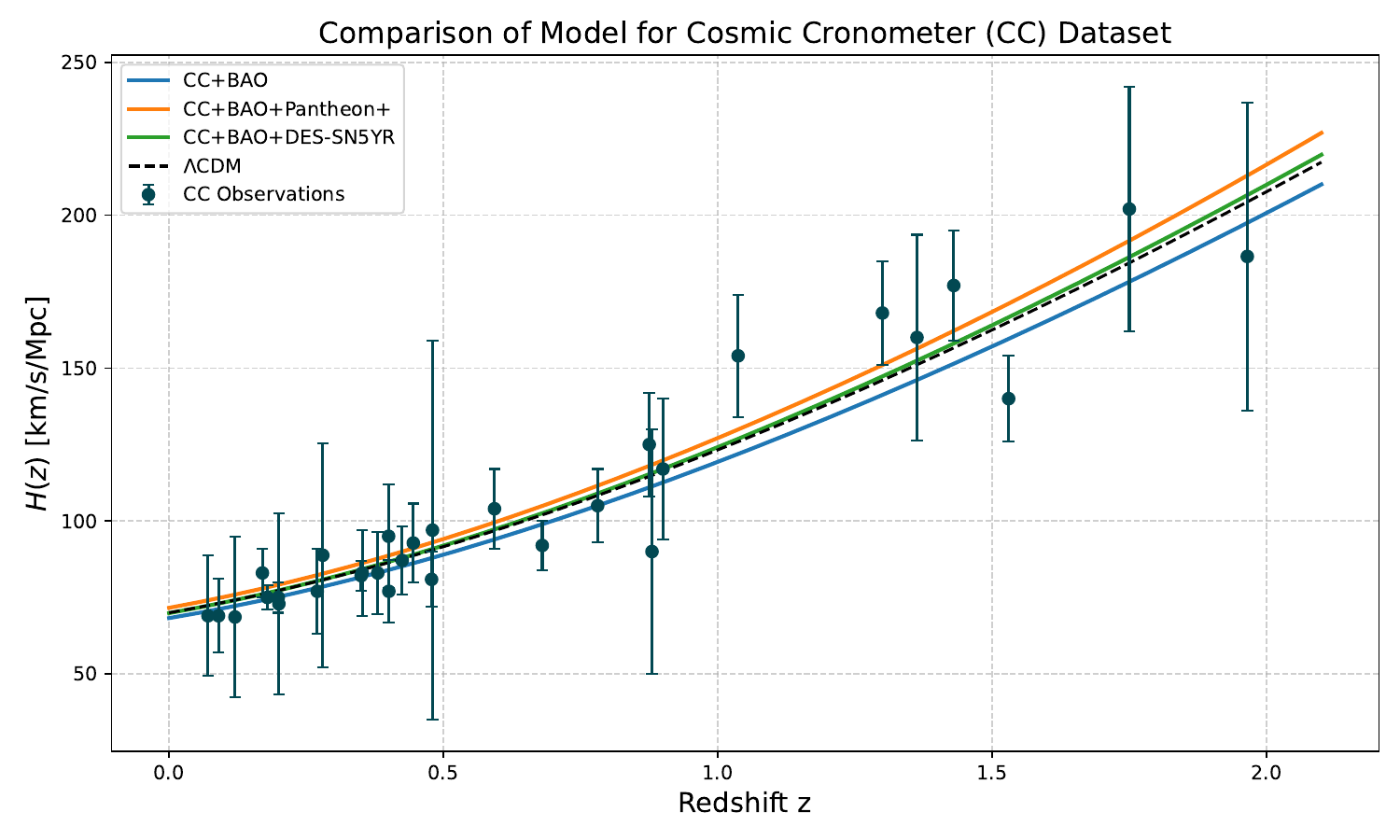}
\caption{Comparison of the reconstructed $H(z)$ for the scalar field model obtained using CC+BAO, CC+BAO+Pantheon$^+$, and CC+BAO+DES-SN5YR datasets. The theoretical curves constrained by these observations are displayed alongside the CC measurements and their respective error bars. For comparison, the $\Lambda$CDM prediction based on best-fit parameters is also displayed.}
\label{Fig2}
\end{figure}

In Figure~\ref{Fig2}, we present a comparison of the theoretical predictions of the model under consideration in this work with the CC observational data, as indicated in the figure caption. The variation of the Hubble parameter $H(z)$ (in km\,s$^{-1}$Mpc$^{-1}$) is illustrated as a function of redshift $z$. The solid curves correspond to the model constraints obtained using different observational combinations, namely CC+BAO (blue), CC+BAO+Pantheon$^+$ (orange), and CC+BAO+DES-SN5YR (green). On the other hand, the black dashed curve represents the predictions from the standard $\Lambda$CDM model. The circular points with error bars represent the CC observational data along with their uncertainties. It can be observed that all model combinations exhibit consistency with the observational $H(z)$ measurements throughout the considered redshift range. Furthermore, the inclusion of additional datasets, such as Pantheon$^+$ and DES-SN5YR, leads to a moderate reduction in the deviation from $\Lambda$CDM and tightens the uncertainty bands. Hence, in general, we can say that the inclusion of additional observations has an overall impact on the fit. This indicates that the proposed model can effectively reproduce the observed late-time acceleration of the universe when constrained with combined datasets.

\subsubsection*{\textbf{Distance Modulus Reconstruction from Pantheon$^+$ and DES-SN5YR Supernovae Data}}
\begin{figure}[htbp]
\centering
\includegraphics[width=1.1\textwidth]{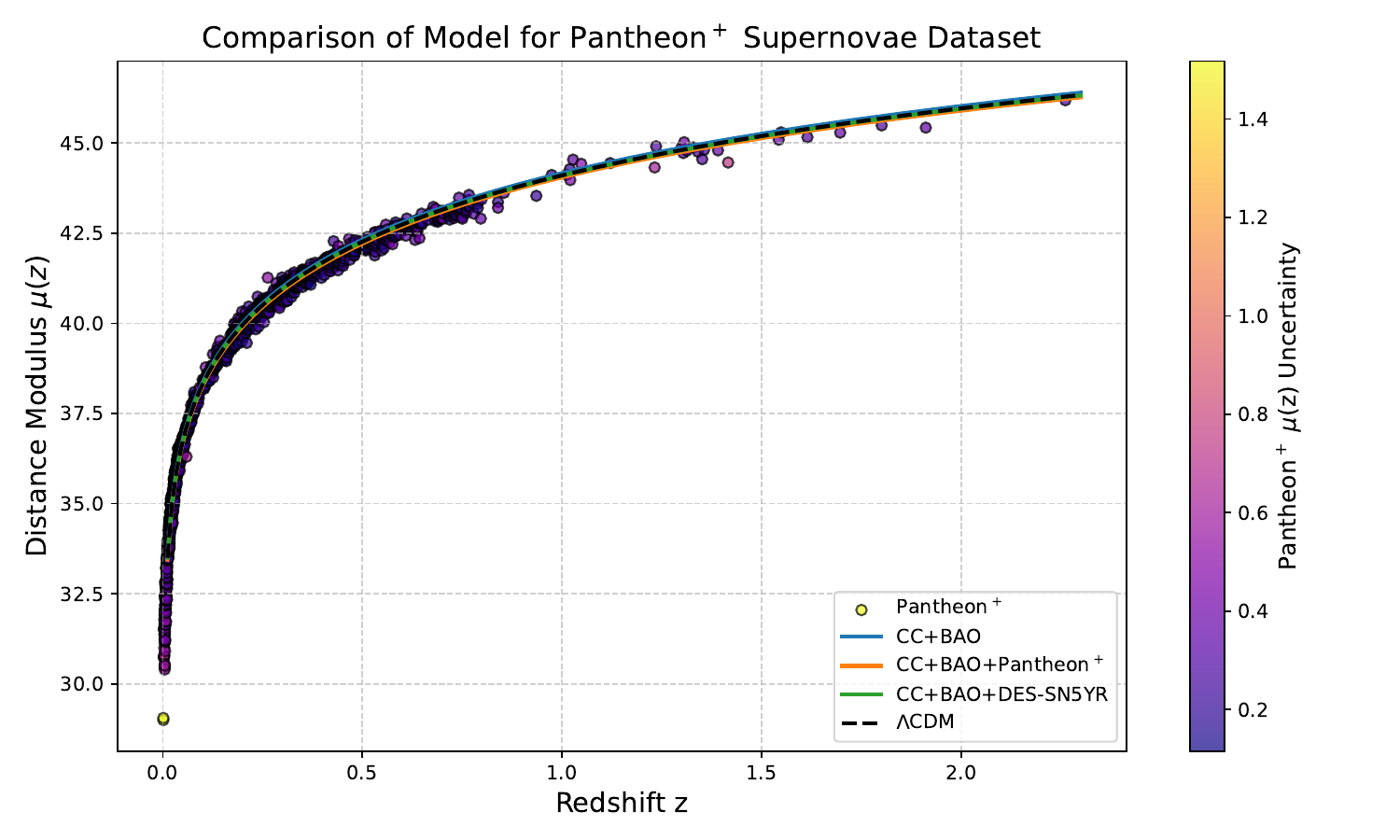}
\caption{Comparison of the reconstructed $\mu(z)$ for the scalar field model obtained using CC+BAO, CC+BAO+Pantheon$^+$, and CC+BAO+DES-SN5YR datasets. The curves are shown together with the Pantheon$^+$ observations. For reference, the $\Lambda$CDM prediction based on the corresponding best-fit parameters is also displayed. The colour code represents the measure of uncertainty.}
\label{Fig3}
\end{figure}

Let us now come to Figure~\ref{Fig3}. This figure illustrates the comparison between the theoretical predictions of the reconstructed model and the observational Pantheon$^+$ dataset in terms of the distance modulus $\mu(z)$ as a function of the redshift $z$. The colored circular points with associated error bars represent the Pantheon$^+$ observational data along with their $1\sigma$ uncertainties. On the right side of the plot, we have shown the color scale that denotes the variation in observational uncertainty of $\mu(z)$. The solid curves in the plot presented in this figure correspond to the theoretical fits for the model constrained with different dataset combinations—CC+BAO (blue), CC+BAO+Pantheon$^+$ (orange), and CC+BAO+DES-SN5YR (green). On the other hand, the dashed black curve indicates the standard $\Lambda$CDM prediction for comparison. It is observed from the figure that all model combinations closely follow the observed trend of SNe Ia. This implies that there is considerable agreement across the entire redshift range up to $ z\simeq 2.3$. The inclusion of additional datasets, such as Pantheon$^+$ and DES-SN5YR, slightly refines the accuracy of the fitting and reduces the dispersion of the residuals. This helps us conclude that the reconstructed model can successfully reproduce the observed late-time cosmic acceleration behavior, consistent with $\Lambda$CDM cosmology.

\begin{figure}[htbp]
\centering
\includegraphics[width=1.1\textwidth]{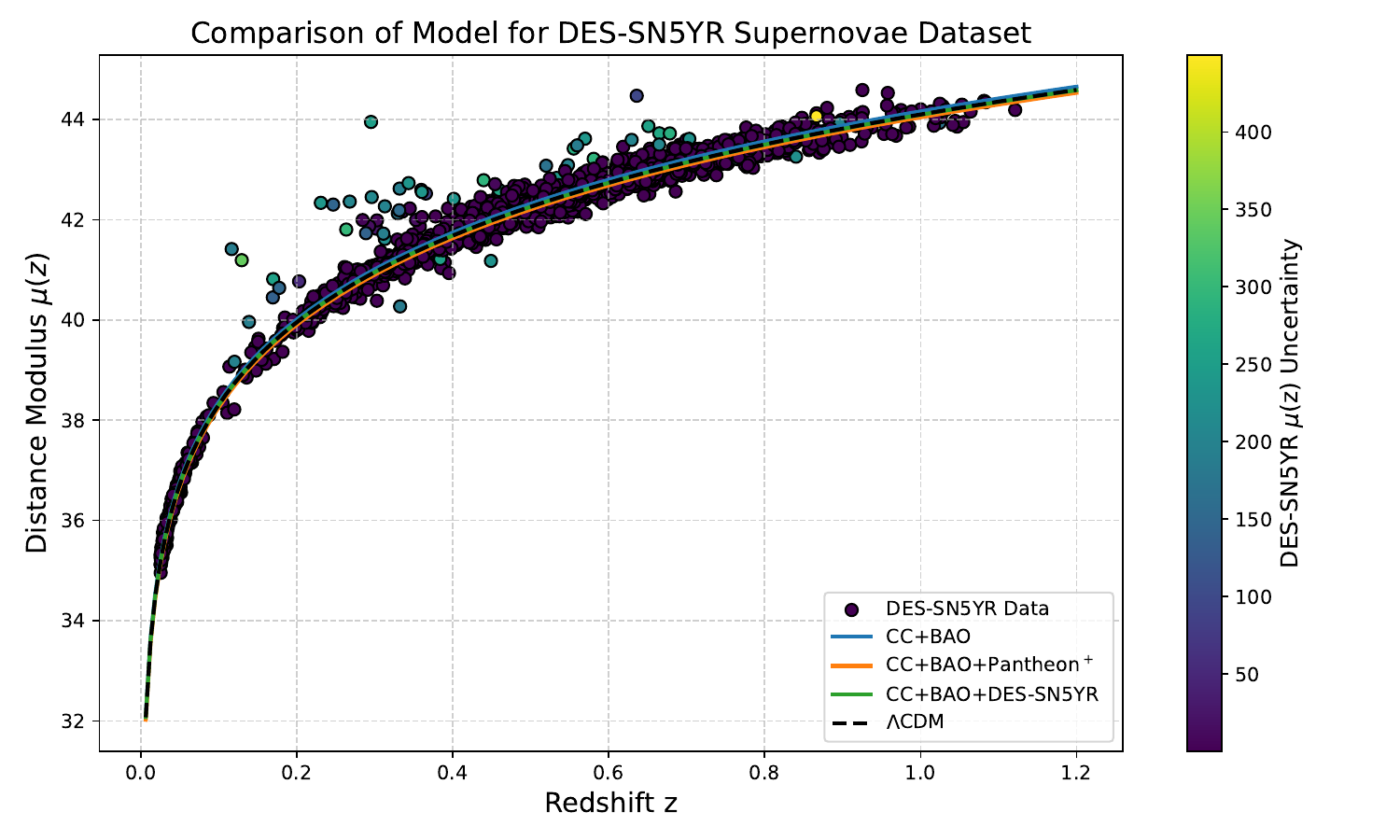}
\caption{Comparison of the reconstructed $\mu(z)$ for the scalar field model obtained using CC+BAO, CC+BAO+Pantheon$^+$, and CC+BAO+DES-SN5YR datasets. The curves are shown together with the DES-SN5YR observations. For reference, the $\Lambda$CDM prediction based on corresponding best-fit parameters is also displayed. The colour code represents the measure of uncertainty.}
\label{Fig4}
\end{figure}

Fig.~\ref{Fig4} depicts the comparison of the model's theoretical prediction of the distance modulus $\mu(z)$ with the DES-SN5YR SNe Ia dataset. The colored circular markers with their corresponding error bars represent the Pantheon$^+$ SNe Ia measurements together with the associated $1\sigma$ uncertainties. A color scale on the right-hand side indicates the variation in the observational error of $\mu(z)$. The solid curves show the model predictions obtained using different combinations of datasets—CC+BAO (blue), CC+BAO+Pantheon$^+$ (orange), and CC+BAO+DES\text{-}SN5YR (green). For reference, the dashed black line depicts the standard $\Lambda$CDM expectation. It is observed that for $z\lesssim 0.1$, the models converge, as expected, due to the uniformity of cosmic expansion in this regime. Apart from this, there is a visibly significant agreement throughout the range of redshift reported in the plot. As redshift increases, we observe minor deviations among different combinations of the datasets, as already elaborated. From this, we understand their sensitivity to the cosmological parameters. This consistency helps us to confirm that the proposed model can successfully reproduce the observed late-time acceleration of the universe while maintaining strong compatibility with standard cosmology.

\subsubsection*{\textbf{Scaled Comoving Angular Diameter Distance and Comparison with BAO Observations}}
\begin{figure}[htbp]
\centering
\includegraphics[width=1\textwidth]{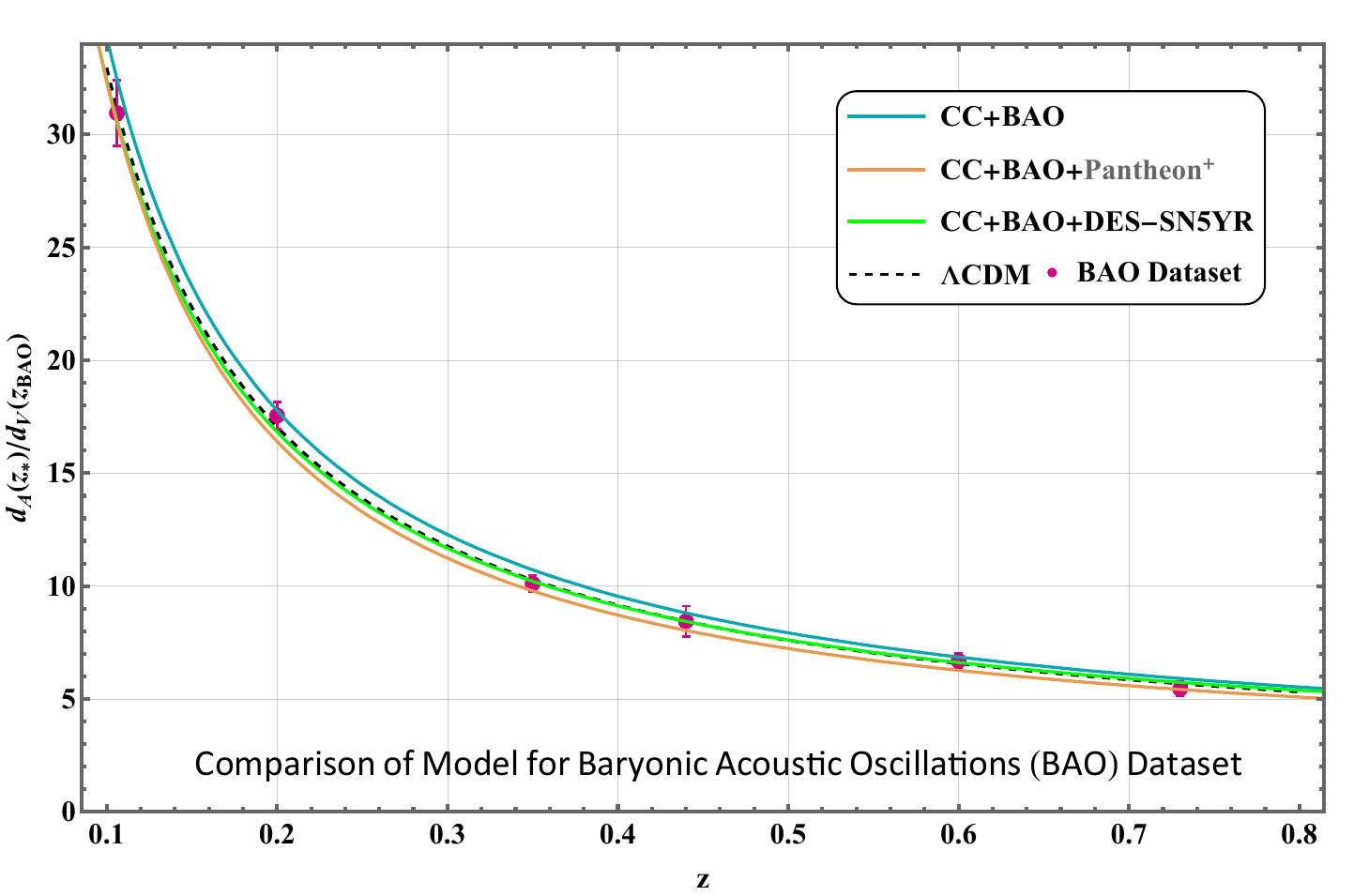}
\caption{Comparison of the scaled comoving angular diameter distance versus redshift z for the scalar field model obtained using CC+BAO, CC+BAO+Pantheon$^+$, and CC+BAO+DES-SN5YR datasets. The model predictions obtained from the observational constraints are shown. The measured BAO data points are displayed with error bars. For reference, the corresponding $\Lambda$CDM curve, constructed using its best-fit parameters, is also included.}
\label{Fig5}
\end{figure}

In Figure~\ref{Fig5}, we have illustrated a comparative study of the scaled comoving angular diameter distance plotted against redshift for three combinations of the observational data. A monotone decreasing behavior of the angular diameter distance is observed with respect to z. Hence, it is understandable that, with the evolution of the universe, the comoving angular distance increases. As we compare the three curves with the $\Lambda$CDM curve, we observe that all the curves are in good alignment with the BAO data, whose error bars are confined to the neighborhood of the curves. Furthermore, the $\Lambda$CDM curve (black dotted) almost entirely coincides with CC+BAO (blue), CC+BAO+Pantheon$^+$ (orange), and CC+BAO+DES-SN5YR (green) datasets as observed from almost coincident behavior OF the green and black dotted curves. This further consolidates the robustness of the model tested against observational datasets. Additionally, the inclusion of DES-SN5YR in the CC+BAO enhances model stability. 

The determination of the present Hubble expansion rate remains one of the central issues in modern cosmology due to the well-known tension between early- and late-universe measurements. The SH0ES Collaboration reported a relatively high value of $H_0 = (73.04 \pm 1.04)\ \text{km s}^{-1}\text{Mpc}^{-1}$ based on the Cepheid-calibrated Type Ia supernova distance ladder \cite{riess2022comprehensive}. 

From our analysis, summarized in Table~\ref{table:1}, the combined dataset CC+BAO+Pantheon$^+$ yields an estimate of $H_0$ that is consistent with the SH0ES determination within the corresponding uncertainties. In contrast, when the Pantheon$^+$ Supernova sample is excluded, as in the cases of the CC+BAO and CC+BAO+DES-SN5YR combinations, the inferred values of $H_0$ deviate from the SH0ES measurement and shift toward comparatively lower values.

For reference, the Planck 2018 analysis of the cosmic microwave background reported a lower value of the Hubble constant, $H_0 = (67.4 \pm 0.5)\ \text{km s}^{-1}\text{Mpc}^{-1}$ \cite{aghanim2020planck}. In our results, the CC+BAO dataset produces an estimate that lies closer to the Planck determination. When the Pantheon$^+$ dataset is incorporated, the reconstructed value of $H_0$ shifts toward the SH0ES measurement, whereas the CC+BAO+DES-SN5YR combination yields an intermediate value located between the Planck and SH0ES limits.

These results suggest that within the exponential quintessence scalar field framework considered here, the inferred Hubble constant depends noticeably on the adopted observational dataset. The model does not completely resolve the $H_0$ tension, but it allows parameter estimates that lie between the early-universe (Planck) and late-universe (SH0ES) determinations, indicating a moderate shift in the inferred expansion rate relative to the standard $\Lambda$CDM expectations.

\subsection{Statistical Model Selection Using Akaike Information Criterion}
In this subsection, we evaluate the statistical viability of the considered cosmological scenarios using information-theoretic model selection criteria. In particular, we employ the Akaike Information Criterion (AIC), which is designed to assess the relative performance of competing models while incorporating a penalty for increasing model complexity. This criterion provides an effective framework for avoiding overfitting by balancing the quality of the fit against the number of model parameters.

Such statistical measures have been widely adopted in cosmological analyses. For instance, in kinematic studies based on Type Ia supernovae and $H(z)$ measurements \citep{benndorf2022determination}, as well as in investigations involving the calibration of gamma-ray burst (GRB) Amati relations using Gaussian Process reconstructed data \citep{han2024detection}, the AIC has often indicated a preference for comparatively simpler cosmological descriptions. This highlights its usefulness in guiding model selection toward statistically efficient and parsimonious representations of observational data.

To examine the statistical performance of the proposed exponential quintessence framework, we compute the AIC value for the model. This criterion helps determine whether the improvement in the goodness of fit justifies the inclusion of additional parameters \cite{cavanaugh2019akaike, mangan2017model}. As discussed in Ref.~\cite{rezaei2021comparison}, the AIC is extensively used in cosmology to compare theoretical models while accounting for the trade-off between descriptive accuracy and parameter economy. Generally, the AIC emphasizes predictive capability and serves as an effective statistical tool for comparing competing cosmological models while accounting for the balance between goodness of fit and the number of free parameters.

In addition, Ref.~\cite{arevalo2017aic} employed this criterion to investigate both linear and nonlinear interaction models between dark matter and dark energy within the framework of general relativity. Using multiple observational datasets, including SNIa, $H(z)$, BAO, and CMB measurements, that work demonstrated how the AIC can be used to distinguish among interacting cosmological models and identify scenarios that may potentially alleviate the cosmic coincidence problem.
The AIC is defined as follows \cite{rezaei2021comparison}:
\begin{align}
\text{AIC} &= \chi^2_{\text{min}} + 2k,
\end{align}
where $k$ denotes the total number of free parameters in the model.

In model comparison, the difference in AIC values between two models, $\Delta\text{AIC}$, provides a quantitative measure of their relative statistical support. Generally, $\Delta\text{AIC} \leq 2$ indicates substantial support for a model, $4 \leq \Delta\text{AIC} \leq 7$ suggests considerably less support, while $\Delta\text{AIC} > 10$ implies that the model is strongly disfavored relative to the reference model \cite{burnham2004multimodel}.

\begin{figure}
    \centering
    \includegraphics[width=0.8\linewidth]{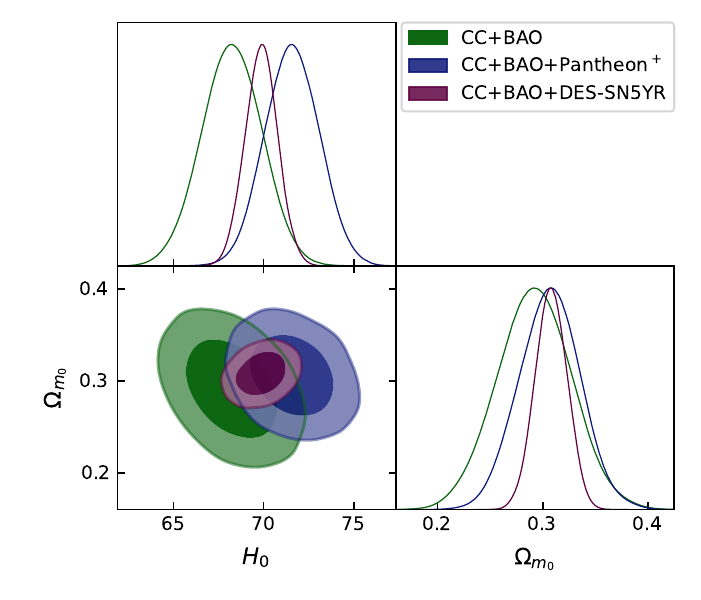}
    \caption{{A two-dimensional confidence contour for the parameters $H_0$ and $\Omega_{m_0}$ of the $\Lambda$CDM model obtained from three different observational dataset combinations: CC+BAO, CC+BAO+Pantheon$^+$, and CC+BAO+DES-SN5YR. The contours represent the $1\sigma$ and $2\sigma$ confidence regions. The plot shows the correlation between the present Hubble parameter and the matter density parameter.}}
    \label{fig:placeholder}
\end{figure}

In the initial stage of our analysis, we present the corner plot of the marginalized posterior distributions for the parameters $H_0$ and $\Omega_{m_0}$ of the $\Lambda$CDM model in Figure~\ref{fig:placeholder} obtained using the CC+BAO (green), CC+BAO+Pantheon$^{+}$ (blue), and CC+BAO+DES-SN5YR (maroon) datasets. The contours correspond to the $1\sigma$ and $2\sigma$ confidence levels. For the CC+BAO dataset, the estimated values of $H_0$ and $\Omega_{m_0}$ remain consistent with the constraints reported by the Planck 2018 results \cite{aghanim2020planck}. It is evident that the inclusion of Supernovae datasets significantly tightens the constraints compared with the CC+BAO case, indicating their strong constraining power on the background expansion history. A mild degeneracy between $H_0$ and $\Omega_{m_0}$ is also observed, which is typical in $\Lambda$CDM analyses. Overall, the parameter estimates remain mutually consistent across the dataset combinations, while the addition of Supernovae data improves the precision of the cosmological constraints. The best-fit values of the cosmological parameters and statistical indicators for the $\Lambda$CDM model corresponding to these datasets are summarized in Table~\ref{table:3}.

\begin{table}[hbt]
\renewcommand\arraystretch{1.5}
\centering 
\begin{tabular}{||c|c|c|c|c||} 
\hline\hline 
~~~Dataset~~~&~~~~~~~ $H_{0}$ ~~~~~~~& ~~~~~~~$\Omega_{m_0}$~~~~~~~ & ~~~~~~~$\chi^2_{min}$~~~~~~&~~~~~$AIC$~~~~~~\\ [0.5ex]  
\hline\hline
CC + BAO & $68.21^{+1.72}_{-1.75}$ &   $0.292^{+0.034}_{-0.036}$ & $19.58$ & $23.58$ \\[0.5ex]
\hline
CC + BAO + Pantheon$^+$ & $71.62^{+1.51}_{-1.49}$ &  $0.307\pm0.029$ & $1638.62$ & $1642.62$\\[0.5ex]
\hline
CC + BAO + DES-SN5YR & $69.87^{+0.92}_{-0.89}$ &  $0.308\pm0.015$ & $1778.43$ & $1782.43$\\[0.5ex]
\hline \hline 
\end{tabular}
\caption{{Constrained parameter values, $\chi^2_{min}$ and AIC for the $\Lambda$CDM model are obtained using three observational datasets: CC + BAO, CC + BAO + Pantheon$^+$, and CC + BAO + DES-SN5YR.}}
\label{table:3} 
\end{table}

{In the subsequent stage of the analysis, we employ the key information criteria: the minimum chi-squared ($\chi^2_{\rm min}$) and the AIC, to perform a statistical comparison between the constrained exponential quintessence model under investigation and the standard $\Lambda$CDM scenario. These metrics are computed for each model using three combinations of observational datasets as already mentioned. The statistical comparison between the exponential quintessence model and the $\Lambda$CDM scenario is summarized in Table~\ref{table:4}. While the $\chi^2$ minimization method provides reliable constraints on model parameters, it does not indicate that the proposed model is statistically favored over $\Lambda$CDM. Although the exponential quintessence model yields slightly smaller minimum chi-square values than the $\Lambda$CDM model for all three dataset combinations, the differences $|\Delta\chi^2_{\min}|$ remain relatively small. This indicates that both models provide a very similar level of agreement with the observational data.}

{To assess whether the improved goodness of fit justifies the additional model parameters, we evaluate the information theory-based model selection criterion, i.e., AIC. The corresponding $\Delta$AIC values are $3.79$, $3.19$, and $2.92$ for the CC+BAO, CC+BAO+Pantheon$^+$, and CC+BAO+DES-SN5YR datasets, respectively. According to the conventional interpretation of $\Delta$AIC, values in the range $2<\Delta\text{AIC}<4$ \cite{burnham2004multimodel} indicate moderate support relative to the reference model. Therefore, although the exponential quintessence model provides a marginally better fit to the data in terms of $\chi^2_{\min}$, the additional model parameters introduce a penalty in the AIC evaluation. As a consequence, the $\Lambda$CDM model remains slightly favored from the perspective of information criteria, while the exponential quintessence scenario still remains statistically competitive with the observational datasets considered here.}

\begin{table}[hbt]
\renewcommand\arraystretch{1.5}
\centering 
\begin{tabular}{||c|c|c|c|c||} 
\hline\hline 
~~~Dataset~~~&~~~~~~~$\chi^2_{min}$~~~~~~&~~~~~$AIC$~~~~~~&~~~~~$\lvert\Delta{\chi^{2}_{min}}\rvert$~~~~~&~~~~~$\Delta{AIC}$~~~~~\\ [0.5ex]  
\hline\hline
CC + BAO & $19.37$ & $27.37$ & $0.21$ & $3.79$ \\[0.5ex]
\hline
CC + BAO + Pantheon$^+$ & $1637.81$ & $1645.81$ & $0.81$ & $3.19$  \\[0.5ex]
\hline
CC + BAO + DES-SN5YR & $1777.35$ & $1785.35$ &  $1.08$ & $2.92$  \\[0.5ex]
\hline \hline 
\end{tabular}
\caption{{$\chi^2_{min}$ and AIC for the observationally-constrained exponential quintessence model, and $\Delta\chi^{2}_{min}$ and $\Delta$AIC (differences in comparison with the $\Lambda$CDM model) using three observational datasets: CC + BAO, CC + BAO + Pantheon$^+$, and CC + BAO + DES-SN5YR.}}
\label{table:4} 
\end{table}

{At this juncture, we note that the Pantheon$^{+}$ and DES-SN5YR compilations represent two of the most precise Type Ia Supernovae datasets currently available \cite{scolnic2022pantheonplus,brout2022pantheonplus,abbott2024dark,vincenzi2024dark}. Since these samples originate from different surveys and employ distinct calibration pipelines, careful treatment of potential systematic uncertainties is necessary when interpreting cosmological constraints \cite{vincenzi2025comparing,dinda2025calibration,efstathiou2025evolving}. In the present analysis, the Pantheon$^{+}$ and DES-SN5YR datasets are not merged into a single Supernova catalog. Instead, they are incorporated separately in the likelihood analysis, with independent likelihood functions constructed for each dataset. This approach avoids possible cross-calibration biases that could arise from combining surveys with different photometric calibration schemes. Consequently, the parameter constraints derived from different dataset combinations (see Table~\ref{table:1}) remain statistically consistent. Moreover, this procedure enables a clearer assessment of the individual impact of each Supernova dataset on the estimation of cosmological parameters.} 

\section{Cosmological Parameters}\label{sec4}
This section is devoted to an insightful discussion of the evolutionary patterns of various cosmological parameters, which necessitate a thorough analysis to comprehend the dynamics of evolution. Since we are assessing candidates for late-time acceleration, it is necessary to focus on the EoS parameter, deceleration parameter, and fractional energy densities associated with the matter and scalar field under consideration. These quantities are directly expressible through the reconstructed Hubble parameter and hence can provide essential diagnostics of the universe's dynamical evolution. A consistent comparison $\Lambda$CDM paradigm is therefore possible from this examination. The following subsections give details of the analysis of different cosmological parameters.

\subsection{Total EoS and Deceleration Parameters, and Fractional Energy Densities}

The total EoS parameter \cite{koussour2023thermodynamical} and the deceleration parameter \cite{yadav2024reconstructing} can be formulated in terms of the Hubble parameter $H(z)$ and the redshift $z$ as follows:
\begin{equation}\label{10:22}
\omega_{\mathrm{tot}} = -1 + \frac{2(1+z)}{3H(z)} \frac{dH(z)}{dz},
\end{equation}
\begin{equation}\label{10:23}
q = -1 + \frac{(1+z)}{H(z)} \frac{dH(z)}{dz}.
\end{equation}
Hence, we can re-express the total EoS and deceleration parameters, respectively, in terms of the dimensionless variables using Eq.~(\ref{10:7}) as
\begin{equation}\label{10:022}
\omega_{\mathrm{tot}}=-1+\frac{2\eta^2+\xi}{h^2},
\end{equation}
\begin{equation}\label{10:023}
q=-1+\frac{3}{2}\left[\frac{2\eta^2+\xi}{h^2}\right].
\end{equation}

\begin{figure}[ht!]
    \centering
    \includegraphics[width=18cm]{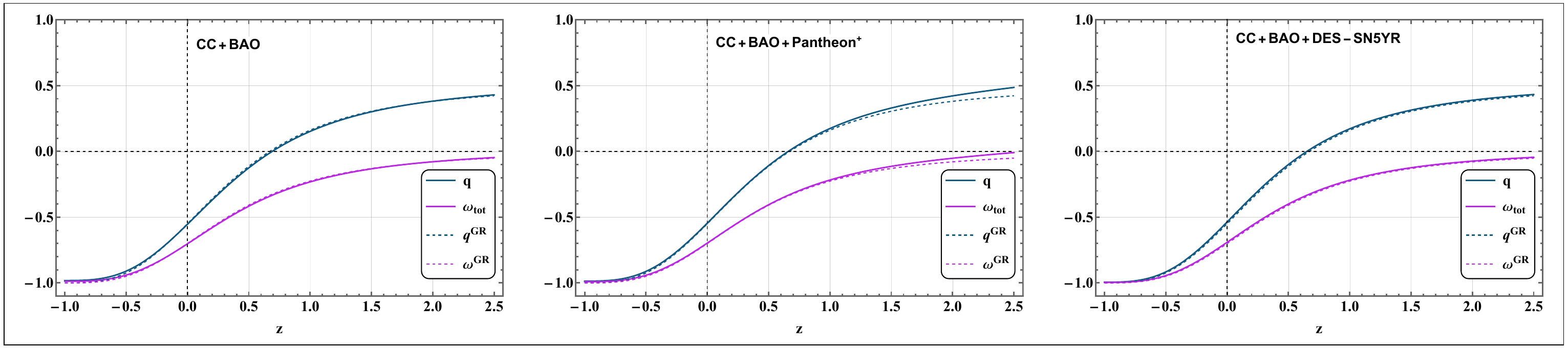}
    \caption{Redshift evolution of the deceleration parameter $q(z)$ and the total EoS parameter $\omega_{\mathrm{tot}}(z)$ for the quintessence model with an exponential potential, compared with the standard $\Lambda$CDM background. Results are shown for three observational combinations: CC+BAO, CC+BAO+Pantheon$^+$, and CC+BAO+DES-SN5YR.}
    \label{Fig7} 
\end{figure}

From the Figure~\ref{Fig7}, we observe that the evolution of the universe through redshift is successfully traced with the model under consideration. At early times (high redshift), the universe behaves as a matter-dominated universe. With the decrease of the redshift towards the current universe, the influence of the scalar field leads to a transition to accelerated expansion. This phase is characterized by a negative deceleration parameter. In comparison with the $\Lambda$CDM reference curves, the quintessence scenario exhibits mild deviations at late times: the total EoS parameter becomes negative, which signifies dark energy domination. However, it remains above the phantom boundary ($\omega_{\mathrm{tot}} > -1$), which is consistent with the canonical nature of the field and the exponential potential adopted.

The impact of different observational combinations is also evident. Adding the Pantheon$^+$ or DES-SN5YR SNe Ia samples to CC+BAO constrains the late-time evolution more tightly. This understandable from the fact that it is producing a narrower range of allowed trajectories and slightly shifting the transition redshift relative to the case CC + BAO only. These dataset-dependent shifts illustrate that the inferred timing and rapidity of the transition from decelerated to accelerated phase is sensitive to the precise SNe Ia calibration and sample composition. On the other hand, the overall qualitative picture consisting of matter domination at early times followed by scalar-field driven acceleration at late times remains robust.

\begin{figure}[ht!]
    \centering
    \includegraphics[width=18cm]{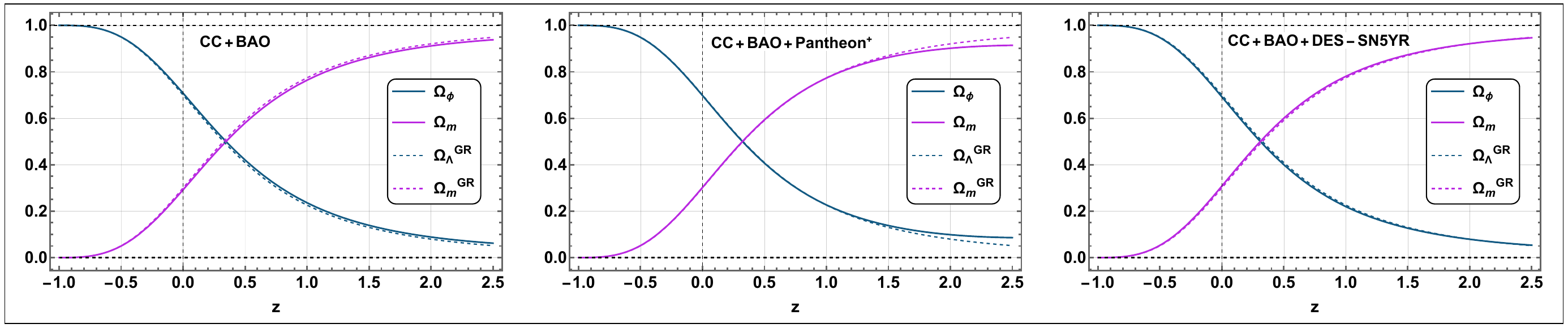}
    \caption{Evolution of the matter density parameter $\Omega_m(z)$ and the quintessence field density parameter $\Omega_\phi(z)$ for the model with an exponential potential, compared with their $\Lambda$CDM counterparts $\Omega_m^{\mathrm{GR}}$ and $\Omega_\Lambda^{\mathrm{GR}}$. The three panels correspond to constraints from CC+BAO, CC+BAO+Pantheon$^+$, and CC+BAO+DES-SN5YR datasets.}
    \label{fig:Density_parameter}
\end{figure}

The evolutionary behavior of the fractional densities of matter and the scalar field is plotted in figure \ref{fig:Density_parameter} as a function of redshift.  At early times ($z \gtrsim 1$), the expansion is governed by non-relativistic matter, consistent with a matter-dominated era. As the universe evolves, the quintessence field gradually becomes dominant, leading to the observed accelerated expansion at late times. As we compare it with $\Lambda$CDM, we observe that the transition is smoother and slightly delayed due to the dynamic nature of the scalar field driven by the exponential potential. The addition of Pantheon$^+$ and DES-SN5YR data refines the constraints and it confirms that the model remains consistent with observations while allowing small deviations that could be tested with future high-precision cosmological data.

\subsection{Statefinder Diagnostics}
To better distinguish between various cosmological scenarios, especially those describing dark energy, the Statefinder diagnostic has been developed as an effective geometric tool for distinguishing between different cosmological scenarios, particularly those involving dark energy. This diagnostic framework utilizes higher-order derivatives of the cosmic scale factor, offering a model-independent method for comparing theoretical models with observational data. The diagnostic pair $(r,s)$ \cite{Sahni_2003_77,alam2003exploring} is directly connected to the scale factor $a(t)$ and hence to the underlying space-time geometry, which justifies its designation as a ``geometric" diagnostic. The Statefinder parameter $r$ can also be reformulated in terms of the Hubble parameter $H(z)$ and the redshift $z$, while the parameter $s$ can be expressed as a function of $r$ and the deceleration parameter $q$. These relations are given by
\begin{equation}
r(z) = 1 + (1+z)^2 \left[ \frac{H''(z)}{H(z)} + \left( \frac{H'(z)}{H(z)} \right)^2 \right]
       - 2(1+z)\frac{H'(z)}{H(z)},
\label{10:24}
\end{equation}
\begin{equation}
s(z) = \frac{r(z) - 1}{3\left[q(z) - \tfrac{1}{2}\right]},
\label{10:25}
\end{equation}
where $H' = \frac{dH}{dz}$ and $H'' = \frac{d^2H}{dz^2}$. These reformulated expressions enable the Statefinder diagnostic to be directly evaluated from the Hubble function reconstructed through observational data. Table~\ref{rs} summarizes the characteristic values of the $(r, s)$ pair for several representative cosmological models, demonstrating that distinct dark energy scenarios occupy different regions in the $(r, s)$ plane.

\begin{table}[h!]
\centering
\renewcommand{\arraystretch}{1.3}
\begin{tabular}{|c|c|c|}
\hline \hline
\textbf{Cosmological Model} & ~~~~~~~~~~~~~~\textbf{$r$} ~~~~~~~~~~~~~~& ~~~~~~~~~~~~~~\textbf{$s$}~~~~~~~~~~~~~~ \\
\hline \hline
$\Lambda$CDM model & $1$ & $0$ \\
\hline
Chaplygin Gas (CG) model & $> 1$ & $< 0$ \\
\hline
Standard Cold Dark Matter (SCDM) model & $1$ & $1$ \\
\hline
Quintessence regime & $< 1$ & $> 0$ \\
\hline
Holographic Dark Energy (HDE) model & $1$ & $\frac{2}{3}$ \\
\hline \hline
\end{tabular}
\caption{Typical values of the Statefinder parameters $(r, s)$ for various cosmological models. Each dark energy scenario occupies a distinct region in the $(r, s)$ plane, allowing for effective discrimination among different models.}
\label{rs}
\end{table}

\begin{figure}[ht!]
    \centering
    \includegraphics[width=18cm]{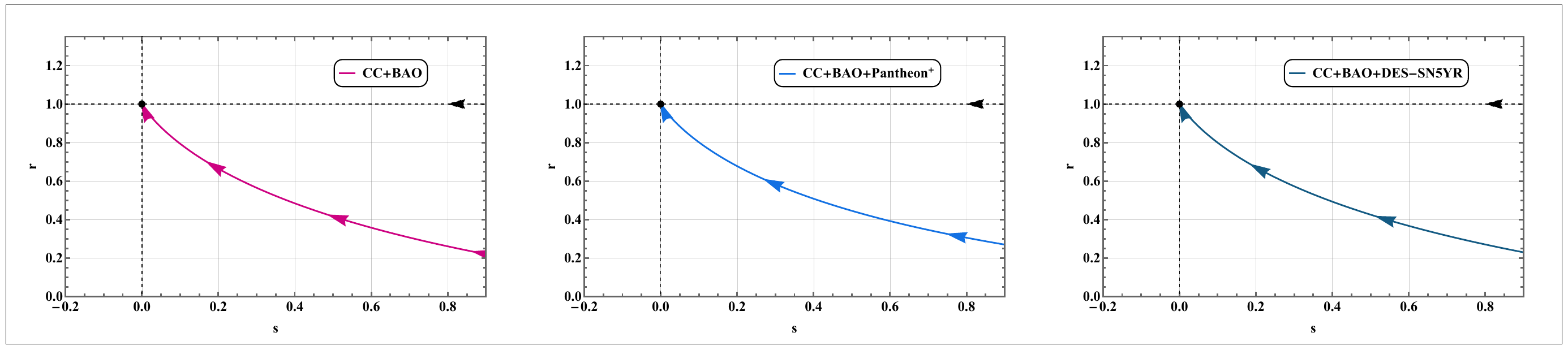}
    \caption{Statefinder trajectories in the $(r, s)$ plane for the quintessence model with an exponential potential, reconstructed using different observational combinations: CC+BAO, CC+BAO+Pantheon$^+$, and CC+BAO+DES-SN5YR.}
    \label{fig:RS_parameter}
\end{figure}

The figure \ref{fig:RS_parameter} illustrates the dynamical behavior of the quintessence model in the Statefinder plane, where $r$ and $s$ are constructed from higher-order derivatives of the scale factor and provide a geometric diagnostic of cosmic acceleration. In all cases, the trajectories evolve from the matter-dominated regime (characterized by $r > 1$, $s < 0$) toward the $\Lambda$CDM fixed point $(1, 0)$ as the universe transitions to accelerated expansion. This evolution demonstrates that the model successfully reproduces late-time acceleration while allowing small deviations from $\Lambda$CDM, primarily governed by the potential slope parameter $\gamma$. The addition of Pantheon$^+$ and DES-SN5YR data leads to a tighter clustering of the trajectories near the point $(1, 0)$, which highlights that the precision of observational constraints is improved and also shows the robustness of the scalar-field framework.

\begin{figure}[ht!]
    \centering
    \includegraphics[width=18cm]{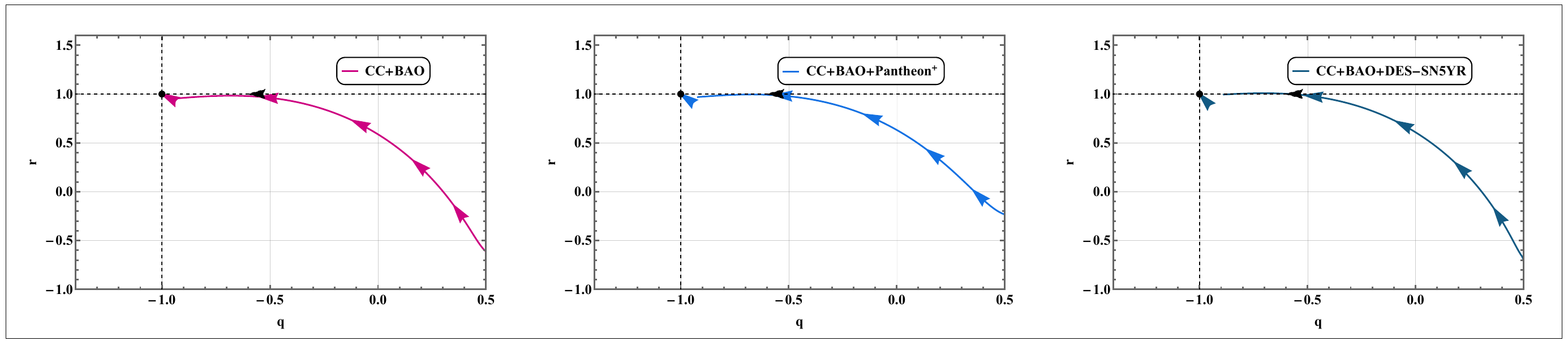}
    \caption{Evolution of the Statefinder trajectory in the $(r, q)$ plane for the quintessence model with an exponential potential, constrained using three different observational datasets: CC+BAO, CC+BAO+Pantheon$^+$, and CC+BAO+DES-SN5YR.}
    \label{fig:RQ_parameter}
\end{figure}

The figure \ref{fig:RQ_parameter} depicts the dynamical evolution of the quintessence model in the $(r, q)$ diagnostic plane, where $r$ is the Statefinder parameter and $q$ is the deceleration parameter. The trajectories show the cosmic evolution from early matter-dominated epochs (characterized by $q > 0$ and $r > 1$) toward the late-time accelerated phase ($q < 0$), eventually approaching the $\Lambda$CDM fixed point $(1, -1)$. This behavior demonstrates that the exponential potential yields a cosmological evolution consistent with observations, exhibiting a smooth transition from deceleration to acceleration. The tighter clustering of the trajectories when SNe Ia data (Pantheon$^+$ or DES-SN5YR) are included to the data implies that these datasets significantly reduce parameter degeneracies and enhance the robustness of the model's predictions.

\subsection{Age Evolution of the Universe}
The present age of the universe, $t_0$, can be determined from the Hubble expansion rate through the integral relation
\begin{equation}
    t_0 = \int_0^{\infty} \frac{dz}{(1+z)H(z)}.
\end{equation}
This expression links the expansion history to the total cosmic age, serving as a crucial self-consistency check for any cosmological model. The obtained value of $t_0$ should remain consistent with current observational estimates, such as those derived from Cosmic Microwave Background (CMB) measurements by the \textit{Planck} mission \cite{aghanim2020planck}, thereby reinforcing the viability of the proposed cosmological framework. 

\begin{figure}[ht!]
    \centering
    \includegraphics[width=0.9\textwidth]{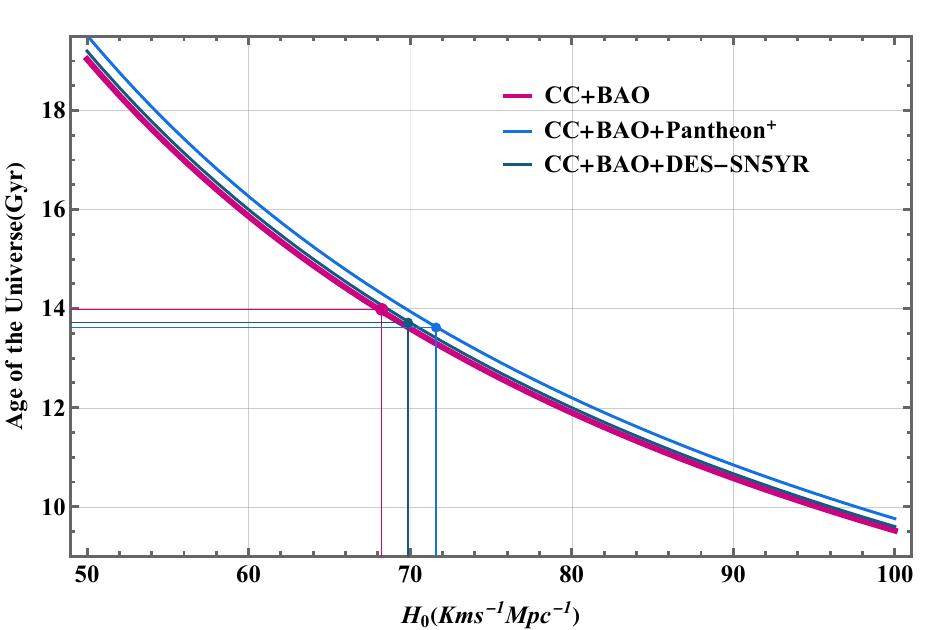}
    \caption{Variation of the estimated age of the universe $t_0$ with the Hubble constant $H_0$ for different observational combinations: CC+BAO, CC+BAO+Pantheon$^+$, and CC+BAO+DES-SN5YR.}
    \label{fig:Age_Universe}
\end{figure}

The figure \ref{fig:Age_Universe} shows the correlation between the present Hubble rate and the estimated age of the universe as obtained from the considered quintessence model. The inverse trend between $t_0$ and $H_0$ reflects the fundamental scaling of cosmic age with the expansion rate, as expected from the integral relation. Among the three observational datasets, the CC+BAO+DES-SN5YR combination yields the most constrained and statistically consistent estimates. The corresponding age aligns well with current cosmological observations, reinforcing the viability of the scalar-field framework in reproducing a realistic cosmic timeline consistent with the \textit{Planck} \cite{aghanim2020planck} 2018 CMB estimate of $t_0 \simeq 13.8\,\mathrm{Gyr}$.

In Table~\ref{table:2}, we list the best-fit present-day value of the essential cosmological parameters associated with the reconstructed scalar field model. The estimates presented in this table are derived from the three observational data combinations mentioned earlier. Tables~\ref {table:1} and~\ref {table:2} illustrate how the inclusion of SNe Ia data points influences cosmological dynamics. 
\begin{table}[htb]
\renewcommand{\arraystretch}{1.5}
\centering
\begin{tabular}{||c|c|c|c||}
\hline\hline
Cosmological Parameter & CC+BAO & CC +BAO+ Pantheon$^+$ & CC + BAO+DES-SN5YR \\
\hline\hline
$q_0$  & $-0.553$ & $-0.539$ & $-0.536$ \\
\hline
$z_{\mathrm{tr}}$ & $0.690$ & $0.649$ & $0.650$ \\
\hline
$w_0$  & $-0.702$ & $-0.692$ & $-0.690$ \\
\hline
$s_0$  & $0.008$ & $0.004$ & $0.002$ \\
\hline
$r_0$  & $0.972$ & $0.985$ & $0.992$ \\
\hline
Age of the universe [Gyr] & $13.985$ & $13.618$ & $13.716$ \\
\hline\hline
\end{tabular}
\caption{Best-fit present-day values of selected cosmological parameters for the reconstructed scaler field model, obtained using three observational datasets: CC+BAO, CC +BAO+ Pantheon$^+$, and CC +BAO+ DES-SN5YR.}
\label{table:2}
\end{table}

\subsection{Energy Conditions and Evolution of the Effective Energy Density}

In this subsection, we analyze the fundamental energy conditions, which represent physical constraints on the combined behavior of energy density and pressure within a given cosmological model. These conditions serve as essential consistency checks, ensuring that the energy-momentum tensor satisfies physically reasonable criteria. In general relativity and its extensions, four principal energy conditions are commonly discussed in the literature: the null energy condition (NEC), weak energy condition (WEC), strong energy condition (SEC), and dominant energy condition (DEC) \cite{yadav2024reconstructing}. Each condition imposes specific inequalities on the energy density $\rho$ and pressure $p$, which can equivalently be expressed in terms of the EoS parameter $\omega = p/\rho$. The standard formulations of these conditions are summarized in Table~\ref{EC}.

\begin{table}[h!]
\centering
\renewcommand{\arraystretch}{1.3}
\begin{tabular}{|c|c|c|}
\hline
\textbf{Condition} & \textbf{Energy Inequality} & \textbf{In Terms of \( \omega = p/\rho \)} \\
\hline
NEC & \( \rho + p \geq 0 \) & \( \omega \geq -1 \) (for \( \rho > 0 \)) \\
\hline
WEC & \( \rho \geq 0, \ \rho + p \geq 0 \) & \( \omega \geq -1 \) \\
\hline
SEC & \( \rho + p \geq 0, \ \rho + 3p \geq 0 \) & \( \omega \geq -\tfrac{1}{3} \) \\
\hline
DEC & \( \rho \geq 0, \ \rho - p \geq 0 \) & \( \omega \leq 1 \) \\
\hline
\end{tabular}
\caption{The four classical energy conditions expressed in terms of the energy density $\rho$, pressure $p$, and the EoS parameter $\omega = p/\rho$.}
\label{EC}
\end{table}

\begin{figure}[ht!]
    \centering
    \includegraphics[width=18cm]{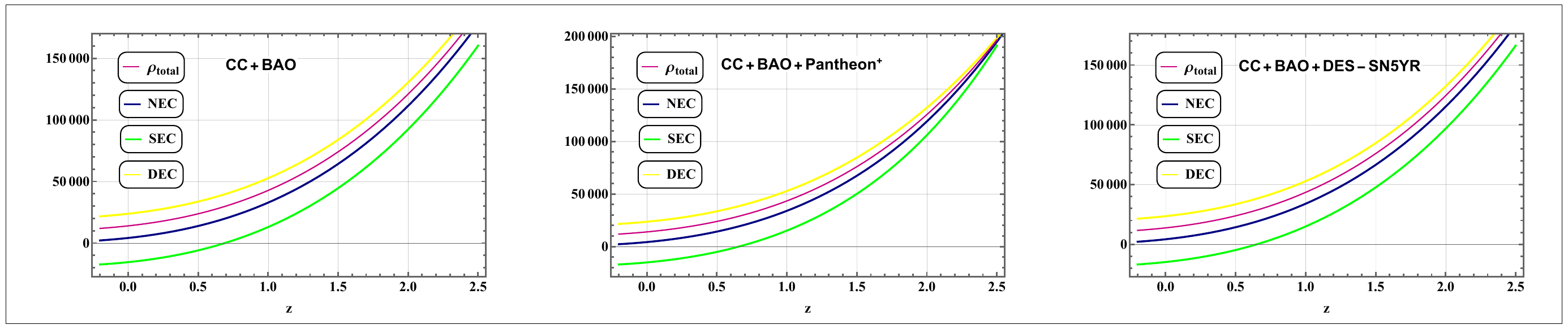}
    \caption{Evolution of the total energy density $\rho_{\text{total}}$ and the energy condition functions (NEC, SEC, DEC) as functions of redshift $z$ for three different observational combinations: CC+BAO, CC+BAO+Pantheon$^+$, and CC+BAO+DES-SN5YR.}
    \label{fig:Energy_Condition}
\end{figure}

The figure \ref{fig:Energy_Condition} presents the redshift evolution of the total effective energy density and the key energy conditions derived for the quintessence model governed by an exponential potential. At high redshifts, all energy conditions remain satisfied, corresponding to a decelerated, matter-dominated era. As the universe evolves, the strong energy condition (SEC) becomes negative, indicating the transition to an accelerated expansion phase driven by the scalar field. The persistent validity of the null (NEC) and dominant (DEC) conditions confirms the physical stability and causal consistency of the model. The results obtained from the CC+BAO+DES-SN5YR combination show the tightest constraints and minimal dispersion, reinforcing the model's robustness under observational scrutiny.

{At very high redshift corresponding to the Big Bang Nucleosynthesis (BBN) epoch ($z\sim10^9$), the universe is expected to be radiation-dominated. In this regime, the Hubble parameter can be approximated as $H(z) \approx H_0\sqrt{\Omega_{r0}}(1+z)^2$.
Consequently, the contribution of the scalar field becomes negligibly small compared with the dominant radiation component. The fractional energy density of the scalar field, $\Omega_{\phi}(z)$, has already been defined in Eq.~(\ref{10:11}). Evaluating this quantity at the BBN epoch using the above approximation shows that $\Omega_{\phi}(z_{\rm BBN})$ approaches zero. Therefore, the model satisfies the primordial nucleosynthesis bound $\Omega_{\phi}(z_{\rm BBN}) < 0.045$, ensuring consistency with standard BBN constraints \cite{bean2001early,steigman2012neutrinos,ade2016planck}.}

\section{Conclusion}\label{sec5}

In this work, we meticulously investigated the cosmological implications of a canonical quintessence scalar field, endowed with an exponential potential, by subjecting it to an exhaustive analysis using the latest available observational datasets, viz., CC, BAO, Pantheon$^+$, and DES-SN5YR.

The exponential potential has long been regarded as an attractive candidate for modeling dynamical dark energy, owing to its natural emergence in higher-dimensional theories, string-inspired constructions, and several modified gravity scenarios. Earlier studies \cite{copeland1998exponential,Heard2002,Shahalam2017,Russo2004,Neupane2004,Kehagias2004} have demonstrated that scalar-field models based on this potential can exhibit scaling regimes and, under suitable conditions, drive accelerated expansion at late times. Building upon these developments, the present work has revisited the exponential potential in light of the most recent high-precision observations. 

Our comprehensive MCMC analysis, summarized in Figure~\ref{Fig1} and Table~\ref{table:1}, clearly shows that including the latest SNe Ia datasets-Pantheon$^+$ and DES-SN5YR significantly improves the precision of the estimated cosmological parameters $\boldsymbol{\Theta} = (H_0, \Omega_{m_0}, \eta_0, \gamma)$. The resulting confidence regions are notably narrower, indicating stronger and more reliable constraints on the model. As expected, we observe a degeneracy between the Hubble constant $H_0$ and the matter density parameter $\Omega_{m_0}$; however, the overall parameter estimates remain consistent and well constrained across all combinations of observational data. The correlation matrices (Figure~\ref{Fig6}) obtained from the CC+BAO, CC+BAO+Pantheon$^+$, and CC+BAO+DES\text{-}SN5YR datasets reveal a consistent parameter dependence structure across all observational combinations. In particular, $H_0$ and $\Omega_{m_0}$ exhibit a strong negative correlation, reflecting the compensatory role these parameters play in shaping the expansion history, while $\eta_0$ and $\gamma$ show similar trends with $\Omega_{m_0}$. Conversely, $H_0$ maintains a positive correlation with both $\eta_0$ and $\gamma$. The persistence of these patterns across datasets underscores the robustness of the inferred parameter relationships within the proposed quintessence framework.

Furthermore, the theoretical predictions obtained from the reconstructed model for the Hubble parameter, $H(z)$, and the distance modulus, $\mu(z)$ (as shown in Figures~\ref{Fig2}, \ref{Fig3}, and \ref{Fig4}), show very good agreement with the CC and SNe Ia observations throughout the entire redshift range studied. This close match provides strong evidence that the model can successfully reproduce the observed late-time acceleration of the universe. The comparison of the scaled comoving angular diameter distance with BAO measurements (Figure~\ref{Fig5}) shows that the model predictions for all dataset combinations closely follow the $\Lambda$CDM curve, with the observational points lying well within the associated uncertainties. The near overlap among the CC+BAO, CC+BAO+Pantheon$^+$, and CC+BAO+DES\text{-}SN5YR curves indicates strong consistency with the standard cosmological model, while the inclusion of DES\text{-}SN5YR data further improves the stability and precision of the reconstructed distance measures. 

The present values of the deceleration parameter, $q_0 \approx -0.54$, and the total EoS parameter, $\omega_{\mathrm{tot}} \approx -0.69$, as listed in Table~\ref{table:2}, clearly confirm that the universe is in an accelerated expansion phase. Importantly, the EoS parameter (Figure~\ref{Fig7}) remains above the phantom divide ($\omega_{\mathrm{tot}} > -1$), which agrees well with the expected behavior of a canonical quintessence field considered in this framework. In addition, the model naturally explains the transition of the universe from an early matter-dominated era to the present scalar-field-driven accelerated phase, occurring smoothly around the transition redshift $z_{\mathrm{tr}} \approx 0.65$. This smooth evolution indicates that the reconstructed scenario is not only theoretically consistent but also observationally viable, making it a strong candidate for describing the late-time cosmic dynamics within the chosen model. The evolution of the fractional energy densities (Figure~\ref{fig:Density_parameter}) indicates the expected transition from a matter-dominated phase at high redshift to late-time acceleration driven by the scalar field. Compared with the $\Lambda$CDM scenario, the transition occurs more smoothly due to the dynamical nature of quintessence. The incorporation of Pantheon$^+$ and DES-SN5YR datasets tightens the constraints and confirms that the model remains observationally viable while allowing mild departures from standard cosmology.

The dynamical nature of the scalar field was further elucidated through the powerful Statefinder diagnostic framework (Figures~\ref{fig:RS_parameter} and \ref{fig:RQ_parameter}). The Statefinder trajectories systematically trace the cosmic evolution from a matter-dominated epoch towards the $\Lambda$CDM fixed point $(1, 0)$ at the present epoch. This confirms that while the model successfully mimics the standard $\Lambda$CDM cosmology at late times, it allows for minor, distinguishing deviations governed by the potential parameter $\gamma$, which future high-precision data can potentially probe.

Finally, the calculated age of the universe (Figure~\ref{fig:Age_Universe}), $t_0$, which was found to be in good alignment with the \textit{Planck} 2018 estimate. This proves the overall self-consistency of our quintessence framework. In the final phase of this study, the model's physical viability was established by examining the Energy Conditions (Figure~\ref{fig:Energy_Condition}). Although it was found that the Strong Energy Condition (SEC) is necessarily violated at late times to sustain cosmic acceleration, the fundamental Null Energy Condition (NEC) and Dominant Energy Condition (DEC) remain satisfied across the entire evolution. The causal consistency and physical stability is confirmed by this result. 

{Our analysis also provides insight into the ongoing $H_0$ tension between early- and late-universe measurements. In particular, the CC+BAO dataset yields an estimate of $H_0$ that is closer to the value inferred from the Planck 2018 CMB observations, $H_0 = (67.4 \pm 0.5)\,\text{km s}^{-1}\text{Mpc}^{-1}$ \cite{aghanim2020planck}. When the Pantheon$^+$ supernova sample is included, the inferred value of $H_0$ shifts toward the higher estimate reported by the SH0ES collaboration, $H_0 = (73.04 \pm 1.04)\,\text{km s}^{-1}\text{Mpc}^{-1}$ \cite{riess2022comprehensive}. In contrast, the CC+BAO+DES-SN5YR dataset yields an intermediate value between the Planck and SH0ES limits. These findings indicate that within the exponential quintessence scalar field framework considered here, the inferred expansion rate depends sensitively on the adopted observational datasets. Although the model does not completely resolve the existing $H_0$ tension, it yields parameter estimates that partially bridge the gap between early- and late-universe determinations. This behavior suggests that dynamical dark energy models with exponential scalar field potentials may offer additional flexibility in describing the late-time expansion history of the universe. Future high-precision cosmological observations will be essential for further assessing the viability of such models in addressing the Hubble tension.}

{In addition to the parameter estimation based on $\chi^2$ minimization, we have also examined the statistical viability of the exponential quintessence model using the Akaike Information Criterion (AIC) (see Tables~\ref{table:3} \& \ref{table:4}). Although the model yields slightly smaller $\chi^2_{\min}$ values compared to the $\Lambda$CDM scenario for all dataset combinations considered, the resulting $\Delta$AIC values lie in the range $2<\Delta\text{AIC}<4$. According to the standard interpretation of the AIC \cite{burnham2004multimodel}, this indicates that the exponential quintessence model is not preferred over the simpler $\Lambda$CDM model, primarily due to the penalty associated with the additional model parameters. Nevertheless, the model remains statistically competitive and provides a viable alternative description of the late-time cosmic expansion consistent with the current observational datasets.}

By combining updated CC measurements, BAO data, and the latest SNe~Ia samples from Pantheon$^+$ and DES-SN5YR, and analyzing them within an MCMC framework, we have placed stringent constraints on the parameter space of a canonical quintessence field governed by the exponential potential. Our findings reveal that the model can reproduce the key features of the observed expansion history while allowing controlled departures from $\Lambda$CDM. The comparison across different dataset combinations further highlights how modern observational probes help refine the behavior of the scalar field and improve the robustness of the derived cosmological dynamics. Overall, the results reaffirm the relevance of exponential-potential quintessence as a viable dynamical alternative to the concordance model, and they underscore the importance of future precision datasets in testing the remaining parameter space of such scenarios.

In summary, the findings of this paper demonstrate that the quintessence model discussed in this study, with its exponential potential, offers a physically viable alternative to the standard $\Lambda$CDM paradigm, serving as a viable candidate for explaining the dark energy sector of the universe. While concluding, let us offer some comments on the outcomes of the current study in relation to the existing plethora of works on scalar field cosmology. Heard et al. \cite{Heard2002} investigated the cosmological dynamics associated with positive and negative potentials, showing that such systems naturally admit scaling solutions that behave like different forms of background fluids. In the context of general relativity, Kehagias et al. \cite{Kehagias2004} considered a spatially flat FLRW universe filled with a minimally coupled scalar field of exponential potential and pressureless baryonic matter, and they found special solutions to possess intervals of acceleration. Russo \cite{Russo2004} provided an exact analytical solution of scalar field cosmologies with exponential potential, illustrating the possibility of transient acceleration. In a relatively recent study \cite{Shahalam2017}, it was demonstrated that a steep exponential potential can support viable late-time acceleration consistent with observational trends. Although these studies established exponential potentials as theoretically viable and attractive candidates to explain late-time acceleration, they essentially relied on dynamical system analysis, exact solutions, or earlier-generation observational data. On the other hand, the present study provides a comprehensive and up-to-date observational test of a canonical quintessence model equipped with an exponential potential by exploring the latest high-precision datasets alongside calculating the age of the universe and testing the energy conditions based on the constrained data. Our study, conducted through MCMC analysis, has tightened constraints on the model parameters by incorporating the Pantheon$^+$ and DES-SN5YR datasets. Overall, while the earlier works established the theoretical viability of the exponential potential quintessence model, the current study builds the connectivity between theory and precision cosmology by demonstrating that such models remain compatible with the latest observational datasets and they can constitute a viable alternative to the standard $\Lambda$CDM paradigm.

\section*{Acknowledgment}
The authors express thankfulness to the anonymous reviewer for the constructive and insightful suggestions to improve the manuscript. The authors also acknowledge Quillbot and Grammarly tools that were used to improve grammar and language.

\section*{Data availability statement}
No new data were generated or analysed in this study. All observational datasets used—Cosmic Chronometers, Baryon Acoustic Oscillation, Pantheon$^+$, and DES-SN5YR Type~Ia Supernovae are publicly available from the corresponding survey repositories. The sources are duly acknowledged in the bibliography.

\section*{References}

\end{document}